\documentclass[aps,prb,floats,amssymb,twocolumn]{revtex4}

\usepackage{graphicx}
\usepackage{docs}
\def\der#1#2{{\partial#1 \over \partial#2}}
\def\be{\begin{equation}}
\def\ee{\end{equation}}
\def\Ho{{\mathcal{H}}_{0}}
\def\Hl{{\mathcal{H}}_{1}}
\def\H{{\mathcal{H}}}

\def\pgs{|G>}
\def\pni{|H>}

\def\sz#1{\sigma^{z}_{#1}}
\def\sx#1{\sigma^{x}_{#1}}
\def\sy#1{\sigma^{y}_{#1}}
\def\ba#1{\begin{array}{#1}}
\def\ea{\end{array}}
\def\r{\right}
\def\l{\left}

\begin{document}

\title{Energy  Correlations In Random Transverse Field Ising Spin
Chains}
\author{{\sc Gil Refael, Daniel S. Fisher}\\{\small \em Dept. of Physics,
Harvard University, Cambridge MA, 02138}}

% -------------------------%

\date{\today}

\begin{abstract}

The end-to-end energy - energy correlations of random transverse-field
quantum Ising spin chains are computed  using a generalization of an
asymptotically exact real-space renormalization group introduced
previously. Away from the critical point, the average energy - energy
correlations decay exponentially with a correlation length that is the
same as that of the spin - spin correlations. The typical
correlations, however, decay exponentially with a characteristic length
proportional to the square root of the primary correlation length.
At the quantum critical point,  the average correlations decay
sub-exponentially as $\overline{C_{L}}\sim e^{-const\cdot L^{1/3}}$,
whereas the typical correlations decay faster, as $\sim
e^{-K\cdot\sqrt{L}}$, with $K$ a random variable with a universal
distribution. The critical energy-energy correlations behave very
similarly to the smallest gap, computed previously; this is explained
in terms of the RG flow and the excitation structure of the chain.  In order to obtain the energy
correlations, an extension of the previously used methods was needed;
here this was carried out via RG transformations that involve a
sequence of unitary transformations.

\end{abstract}

\maketitle

% -----------------------introduction%

\section{Introduction}

The random transverse field Ising model is the prototypical model of a
quantum system with quenched randomness.  It undergoes a novel quantum
phase transition at zero-temperature, and over a wide range of
parameters, exhibits unusual low temperature behavior.  In one
dimension, many of the low energy properties have been found exactly,
initially by McCoy and Wu \cite{mccoy-wu1,mccoy-wu2} and in much detail
recently. \cite{DSF98,DSF95,sachdevsenthil} The behavior in higher
dimensions has been shown to be in the same general class as the one
dimensional system. \cite{Motrunich00, two-dim-numerics}

In addition to its theoretical interest,  models with similar
behavior have been argued to be relevant for the low temperature
properties of heavy fermion materials, with randomness and proximity
to a quantum critical point playing  key roles in producing
non-fermi-liquid behavior. \cite{castroneto}

In this paper we study the end-to-end energy and transverse spin
correlations of long but finite random transverse field Ising chains. 
%The expectation values of
%these correlations provide us with interesting information about the character of the ground state. It provides a
%direct measure of the change in the ground state due to a small change in the strength of a single coupling. 
The average energy and transverse spin correlations  are found to have
rather different behavior than the order parameter
correlations. Computing them introduces new difficulties which
compel us to further develop the general renormalization group (RG)
structure. The  formulation presented here should be useful for other
random quantum systems.

The organization of this paper is as follows; in the remainder of this section
we review some aspects of the random Ising model and introduce the quantities of interest. In
Sec. \ref{formalism} we develop and apply a unitary-transformation approach to the 
real-space RG.  In Sec. \ref{sx-sx}
and \ref{szsz-sx}  Laplace-transforms of the
energy and transverse spin correlations are derived and used in
Sec. \ref{results} to obtain the average and typical correlations, and
more general information about the distributions.  Finally,  Sec.
\ref{conclusions}  presents conclusions.  Some technical details are
confined to an Appendix.

\subsection{Random Transverse Field Ising Model}

\begin{figure}
\includegraphics[width=7cm]{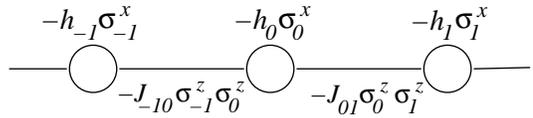}
\caption{The Hamiltonian of the transverse field Ising  model. Each
site is  a spin-1/2 that interacts via Ising exchange  with its
nearest  neighbors and can be flipped by  the  local $x$-magnetic
field.  \label{Hfig}} \end{figure}

The Hamiltonian of the random quantum Ising model is 
\be
\H=-\sum\limits_{i}\left(J_{i\,i+1}\sz{i}\sz{i+1}+h_{i}\sx{i}\right) 
\label{eq:tfisc} 
\ee 
with each site having two states, $\sz{}=\pm1$,  with quantum
fluctuations between them caused by the transverse, $\sx{}$,
fields. The system is illustrated in Fig. \ref{Hfig}.  Note that
there are no magnetic fields in the $z$-direction, so that there is a
global symmetry of inversion about the $xy$ spin plane. The presence
of $z$ fields  would break this symmetry and change the low energy
physics radically.

The quantum Ising model exhibits a quantum phase transition in its ground state when the 
nearest neighbor interaction and the transverse field are of comparable
strength. In a non-random model this occurs when $J=h$.  
In a random system, where the $J$'s and the $h$'s are drawn independently from some  distributions, the transition occurs when 
$\overline{\log h}=\overline{\log J}$, where the over-bars denote averaging over the randomness. A convenient parametrization of the proximity to the transition is   
\be
\delta\equiv\frac{\overline{\log h_I}-\overline{\log J_I}}{var(\log h)+var(\log J)}
\label{eq2}
\ee
with $\delta$ positive in the disordered phase.

\subsection{Real Space RG} A powerful route to  analytic information
on this system is a real space --- or energy space --- RG method that
is a generalization developed by one of us, \cite{DSF95} of an RG
introduced by Ma, Dasgupta and Hu. \cite{MaDas1979,MaDas1980} The real space RG is
carried out by decimating the term in the Hamiltonian --- a site
($h\sx{}$), or a bond ($J\sz{}\sz{}$) --- with the strongest
interaction;  second order perturbation theory results in new effective
couplings. In the case of decimating a bond, a cluster forms with a
renormalized transverse field; in the case of decimating a spin, a new
Ising interaction that couples its two neighbors forms. The
Hamiltonian preserves its form, with effective bonds coupling spin clusters, and the energy scale --- the maximum remaining
coupling ---  reduced.  The effective bond strengths and lengths,
as well as the effective transverse fields on the clusters of spins
and their moments can  be computed.

As the energy scale is systematically reduced, the distributions of
the effective couplings become very broad for small $\delta$.
Concomitantly, the averages of many quantities in the ground state are 
determined by rare tails of their distribution. In particular, the average order parameter correlations,  
\be
\overline{C^{zz}(n,n+r)}\equiv\overline{<\sz{n}\sz{n+r}>},
\ee
at large separations $r$, are dominated by the rare pairs  of spins that are not decimated until they join together into the same spin cluster. \cite{DSF95}

\subsection{Logarithmic Energy Scaling} \label{spinintro}

Many properties of the random quantum Ising model can be understood in terms of the scaling behavior of the cluster sizes, bond lengths, and coupling strengths with the energy scale and the deviation from criticality, $\delta$. 

At the critical point, the distributions of $h_{i}$ and $J_{i\,i+1}$
become infinitely broad as the energy scale, $\Omega$, approaches zero;
the random quantum critical point is thus an  {\it infinite
  randomness} fixed point. \cite{DSF94,DSFstatphys,Motrunich00} At this fixed point the
distribution of cluster and bond lengths, the logarithms of the
interactions in units of $\Omega$,
\be
\ba{c}
\zeta\equiv\log\frac{\Omega}{J},\vspace{2mm}\\ 
\beta\equiv\log\frac{\Omega}{h},
\ea
\ee
and the deviation from criticality, $\delta$, all scale with the logarithm of the energy scale,
\be
\Gamma\equiv\log\frac{\Omega_I}{\Omega}\ .
\ee
Here $\Omega_I$ is the initial energy scale set by the strongest couplings, and $\Omega$ is the magnitude of the largest remaining couplings  after the stronger ones have been decimated.  The scaling can equivalently be given  in terms of a length scale, $\ell$ --- for example the length of an effective bond --- the scaling of log-energies at fixed $\ell$ is of the form
\be 
\ba{c} 
\zeta=z\sqrt{l},\vspace{2mm}\\ 
\Gamma=\gamma\sqrt{l}, 
\ea 
\ee 
where $\gamma$ and $z$ are scale invariant random variables. 

Various  basic results follow directly from this scaling. In particular, the linear number density of remaining spin-clusters at scale $\Gamma$ is 
\be 
n\approx\frac{n_0}{\Gamma^2},
\ee   
with $n_0$ a non-universal prefactor inversely proportional to the original bond lengths.
 
An example of the scaling of log-energies with length is the gap,
$E_1-E_0$, between the ground state and the first excited state of
long finite chains: Analytic and numerical results show that near
the critical point the logarithm of the gap is broadly distributed
on the scale $\sqrt{L}$. Indeed, the distribution of $-\ln
(E_1-E_0)/\sqrt{L}$ attains a universal scaling form in the large
$L$, small $\delta$ limit.\cite{DSF98} From the RG structure, this
can be seen by noting that the gap is the energy scale of the chain
when it has only one remaining cluster --- and thus only one unfrozen
degree of freedom. Therefore the gap is approximately $\Omega_L \sim
e^{-\Gamma_L}$ with $\Gamma_L\propto\sqrt{L}$ the sample-specific
scale at which this last cluster is decimated.

In long finite chains of length $L$, the end-to-end 
spin correlations $<\sz{0}\sz{L}>$ are a useful probe of the long length scale ordering tendencies. The distributions and moments of these can be calculated exactly in the asymptotic limit of long chains and small $\delta$; these  compare well with numerical results. \cite{DSF98} 
The distributions  can be expressed usefully in terms of 
 \be
 \Lambda_z\equiv-\log C^{zz}(0,L) \ .
 \ee
This logarithm of the correlations  scales with $\sqrt{L}$ at the critical point, with a broad distribution on the same scale. The {\it average} correlations, however, decay much more slowly: only as
$\overline{C^{zz}}\propto\frac{1}{L}$.

\subsection{Ordered and Disordered Phases}

When $\delta$ is non-zero but small, there are  two scaling regimes. At early stages of the  decimation process, clusters and interactions are not
``aware'' of being non critical. In this regime the critical scaling
holds. At longer scales, however, there
is a crossover to an off-critical regime. The crossover occurs when the typical cluster sizes and bond
lengths  are of  order of the {\it correlation length}
\be
\xi \approx \frac{1}{\delta^2},
\ee
and the log-energy scale is of order 
\be
\Gamma_\times\sim\frac{1}{\delta} \ .
\ee
At scales larger than $\xi$, the behavior is characteristic of one of the two zero-temperature {\it phases} and thus depends on the sign of $\delta$. 

At low energies in the ordered and disordered phases, the scaling between energy and length scales is different from that at the critical point. For small $\delta$,
in both phases 
\be
\Omega\sim \ell^{-z(\delta)}
\ee
with the effective dynamical exponent,
\be
z\approx \frac{1}{|\delta|}
\ee
near the critical point.

The distributions of the log-interactions also change form. In the
disordered phase, the distribution of the effective fields does not
continue to broaden and $\beta \sim
\Gamma_\times\sim\frac{1}{\delta}$. But the effective bonds become
longer and longer and weaker and weaker with the distribution of the
$\ln J$'s broadening rapidly, with typical $\zeta\sim\Gamma$.
In the disordered phase,  the {\it average} order parameter
correlations decay exponentially with the correlation length
$\xi\approx \frac{1}{\delta^2}$.
Nevertheless,  the typical correlations decay much faster; for
example, end-to-end correlations of almost all samples  decay as
$e^{-2\delta L}$. \cite{shankar-murthy} More precisely, as
$L\rightarrow\infty$, the distribution of of the scaled
log-correlation function, $\Lambda_z/L$, approaches a delta function
peaked at $\Lambda_z/L=2\delta$.  The average correlations are thus
dominated by exponentially rare samples that happen to have
anomalously strong exchanges and/or anomalously weak random fields.

In the ordered phase, $\delta<0$, at low energy scales the clusters become
bigger and bigger, eventually encompassing the whole system.  The
transverse fields on these clusters --- the gap between the symmetric
and anti-symmetric combinations of their two ``ordered" states ---
concomitantly continues to become more and more broadly
distributed. The remaining bonds, on the other hand, stay relatively
short and their distribution does not continue to broaden. These fields
and bonds  thus play opposite roles in the two phases; as discussed
below, this is a general consequence of duality.

%\be 
%\ba{c}
%\begin{array}{cc} 
%\beta=\frac{b}{|\delta|} & \zeta=\frac{z}{|\delta|} 
%\end{array} \vspace{2mm}\\
%\Gamma=\frac{\gamma}{|\delta|}\vspace{2mm}\\ 
%l\delta^2=s, 
%\ea 
%\ee 
%where $z,\,b,\,\gamma,\,s$ the the scale-invariant
%parameters. The cluster size, however, also depends on the the energy scale; the scaling in this case is $l\sim
%e^{\gamma}$, and the inverse of the length becomes an irrelevant parameter.

\subsection{Duality \label{intdual}}
 
As for non-random classical and quantum Ising models, there is a dual
description of the random quantum chain in terms of bond
variables. Instead of using the states  $|\uparrow>,\,|\downarrow>$,
on each site, one can use the states of the {\it bonds}. This is done
by assigning  $|+>,\,|->$ to the bond if the two spins surrounding the
bonds are $|\uparrow\uparrow>,\, |\downarrow\downarrow>$ or
$|\downarrow\uparrow>,\,|\uparrow\downarrow>$ respectively. These are
domain-wall variables. In the new Hilbert space,  if we choose the
quantization axis to be $x$ rather than $z$, the Hamiltonian has the
same form, but with $h$ and $J$ exchanged. The duality is summarized in
the following table: 
\be 
\ba{ccc} 
\ba{c} 
\delta\vspace{2mm}\\
J_{n\,n+1}\vspace{2mm}\\ 
h_n\vspace{2mm}\\ 
\sx{n}\vspace{2mm}\\
\sz{n}\sz{n+1}\vspace{2mm}\\ 
\ea & \Rightarrow & \ba{c}
-\delta\vspace{2mm}\\ 
h_n\vspace{2mm}\\ J_{n\,n+1}\vspace{2mm}\\
\sz{n}\sz{n+1}\vspace{2mm}\\ 
\sx{n}.  
\ea 
\ea 
\ee 
If the $\delta$ dependence of the distributions of the random couplings has the form
$\rho_J(J=X,\delta)=\rho_h(h=X,-\delta)$, the random model is self dual
with $\delta\Rightarrow -\delta$. More generally, it will not be
exactly self dual. But from the definition of $\delta$ in terms of
$\rho_J$ and $\rho_h$ (Eq. \ref{eq2}), and the universality  at low energy scales (in
particular, of the distributions of the effective couplings), we expect
that the asymptotic behavior at low energies and small $\delta$ will
indeed be self-dual for any well-behaved distributions.

\begin{figure}
\includegraphics[width=7.5cm]{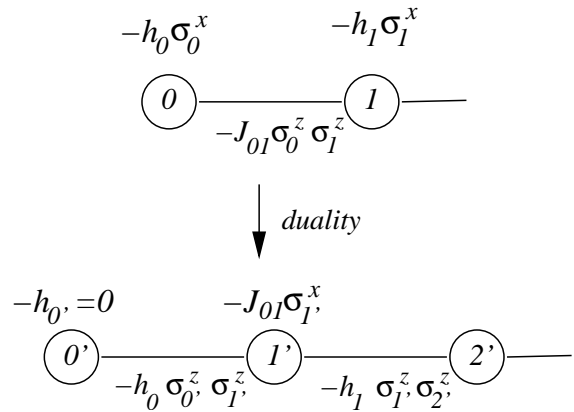}
\caption{A chain that terminates with the site 0 is dual to a chain
  that terminates with a site 0',which experiences no transverse field. The bond
  operator $-h_{0}\sz{0'}\sz{1'}$ is then the edge energy operator
  of the dual chain. \label{endfig}}
\end{figure}

As we are interested in end-to-end correlations of finite chains, we must consider what
happens to the ends of a chain under the duality transformation.
Let $-h_0\sx{0}$ be the energy operator on the left end site. Under
duality, this site will be mapped into a bond with corresponding
energy operator $-h_0\sz{0'}\sz{1'}$ (see Fig. \ref{endfig}). The bond in the dual chain implies a new site 0' that
corresponds to the domain wall variable to the left of the first spin
in the original chain. This new site, 0', thus carries the information
about an arbitrarily fixed boundary condition, e.g., $|\uparrow>$,
with respect to which the leftmost domain wall, and hence the original
end spin $\sz{0}$, is defined; it thus cannot be entirely
forgotten. But in the dual Hamiltonian, the operator $\sx{0'}$ does not
appear, so that the dual end transverse field is zero, and $\sz{0'}$ is
time independent. The same is true at the other end, where the extra
spin is needed for the original $\sx{L}$ to be defined.
Superficially, the dual chain appears to have one extra degree of
freedom associated with each end. But the orientation of the dual
edge-spins is entirely a convention, and therefore the additional
degrees of freedom have no effect.

%The
%time-independence of the extra dual spin at the other end of the chain
%is equivalent to the symmetry under flipping all of the spins in the
%original chain: it parametrizes states as even or odd under this
%global spin flip. 

Note that in the special case in which the original chain starts with
a spin to which  no transverse field is applied, the duality
transformation yields a chain with the first bond turning into a
site.  More generally, any site in the original chain on which there
is  no transverse field corresponds under duality to a break in the
chain that divides it into two uncoupled parts.  In the original
variables, there are concomitantly two disconnected subspaces in which
the spin that cannot flip has $\sigma_z=\pm 1$ respectively. The parts
of the chain to the left and to the right of this spin are thus
independent of eachother.

\subsection{Energy-Energy Correlations}

Our goal in this paper is to understand the energy density (E-E)
correlations of the random chain. These would be particularly
interesting at non-equal times, as they would then yield information
on the transport of energy which is the only locally conserved
quantity in this system.   Unfortunately, in the bulk of the chain
correlations are very hard to calculate for reasons discussed in
ref. \onlinecite{DSF95}. Therefore we study the somewhat simpler but closely
related quantities:  the end-to-end correlations of the  energy
density in finite chains, restricting our analysis to equal time
correlations.

Since the Hamiltonian involves two kinds of terms, $J\sz{0}\sz{1}$ and
$h\sx{}$, to obtain the E-E correlations we need to calculate three
quantities, $<J_{01}\sz{0}\sz{1}h_{L}\sx{L}>$, $<h_0\sx{0}h_L\sx{L}>$
and $<J_{01}\sz{0}\sz{1}J_{L-1\,L}\sz{L-1}\sz{L}>$. However, the
duality transformation simplifies matters, since it maps $h_0\sx{0}h_L\sx{L}$ at $\delta$  to
$J_{01}\sz{0}\sz{1}J_{L-1\,L}\sz{L-1}\sz{L}$ at $-\delta$, so we only
need to compute one of these quantities. Also, the mixed
correlation function is dual to itself, therefore
the distribution of $<J_{01}\sz{0}\sz{1}h_{L}\sx{L}>$ must depend on $|\delta|$, and be
the same in the two phases. Note that a related single-end quantity,
the imaginary time correlation function $<\sx{0}(0)\sx{0}(\tau)>$,
was considered by Igl\'{o}i, Juhasz and Rieger.\cite{igloi}

The calculation of the quantities of interest requires an extension of the
methods used so far. The primary reason for this is that the energy
correlations are dominated by third order perturbative effects at each
stage of the decimation, in contrast to the spin correlations, which
are controlled by second order perturbative effects.

To be able to carry out  higher order RG calculations, we develop an
approach to the decimation in terms of  unitary transformations; this
allows one to  follow precisely how operators of interest (such as
$\sx{}$) evolve  during the RG process. This approach thereby gives a
systematic way to deal with higher orders perturbative effects even in
problems previously analyzed using second order perturbation theory. \cite{DSF98} The
unitary-transformations method is developed in the next section (Sec.\ref{formalism}).

% ----------------------------------formalism%

\section{Unitary Transformation Renormalization Group  \label{formalism}}

In this section we develop a perturbative scheme based on unitary transformations
that will allow us to separate the various parts of the Hamiltonian
and successively simplify the wave functions of the many-body system
to a hierarchical product of simple spin wave
functions. Simultaneously, we must keep track of the original
operators in order to eventually compute their ground state
correlations.

We begin with the first stage of decimation by constructing the
eigenfunction of the highest energy part of the Hamiltonian and
transforming it to take into account the low energy parts
perturbatively. Specifically, the transformation gets rid of the
off-diagonal parts that connect states with large energy differences
between them. For the Ising chain, this can be done while preserving
the {\it form} of the Hamiltonian.

Given a Hamiltonian, $\H$, and a many-body ground state wave function, $\pgs$, with 
\be
\H\pgs=E_{G}\pgs.
\ee
we can generally make a unitary transformation with a hermitian operator $S$ and write:
\be
\ba{c}
e^{iS}\H e^{-iS}e^{iS}\pgs=e^{iS}\pgs,\vspace{2mm}\\  \label{Stransform}
\ba{cc}
\H _{eff}=e^{iS}\H e^{-iS} & \qquad {\rm and}   \ \ \ \ e^{iS}\pgs=\pni
\ea
\ea
\ee
with the goal to make $\pgs$ close to a product of simple wave functions. Such transformations can be used to eliminate --- or separate --- low energy parts in the Hamiltonian. 

Let $\H=\Ho+V$, where $\Ho$ is the high energy  part of $\H$ and $V$
is the remaining low energy parts. The effective Hamiltonian is then
\be
\ba{c}
\H_{eff}=\Ho\vspace{2mm}\\
+V+i[S,\Ho]+i[S,V]+\frac{i^{2}}{2!}[S,[S,\Ho]]+O(S^3). \label{eq:heff}
\ea
\ee
If we are able to choose $S$ so that
\be
V+i[S,\Ho]=0 \label{eq:comm},
\ee
then $\H_{eff}$ will contain no first order terms. The second order corrections to $\Ho$ give rise to  effective interactions. We may now solve for the ground state of $\H_{eff}$ and hence the original $\H$:
\be
\ba{c}
\H_{eff}\pni=E_{G}\pni\vspace{2mm}\\
\pgs=e^{-iS}\pni.
\ea
\ee
Iterating this process separates the higher energy parts of the
Hamiltonian from the lower energy parts. At each stage, the effective
higher energy parts can be simply diagonalized. The remaining
non-diagonalized Hamiltonian only has pieces with energy much lower
then the gap of the high energy section, which was just
diagonalized. The ground-state wave function is then constructed
perturbatively from the wave-function $\pni$, which is a hierarchical wave-function simply
expressible in terms of the ground states of the high energy parts of the
sequence of $H_{eff}$'s. Each term in the hierarchy will be a spin-cluster
pointing in the direction of the transverse field.
For an example see Eq. (\ref{pniex}) below.

Note that this method is related to the flow equation approach for interacting quantum problems  developed  by Kehrein and Wegner. \cite{wegner,stefan}

\subsection{Unitary RG for Transverse Field Ising Chain}

%-----------------------example of h-dec%

\begin{figure}
\includegraphics[width=6cm]{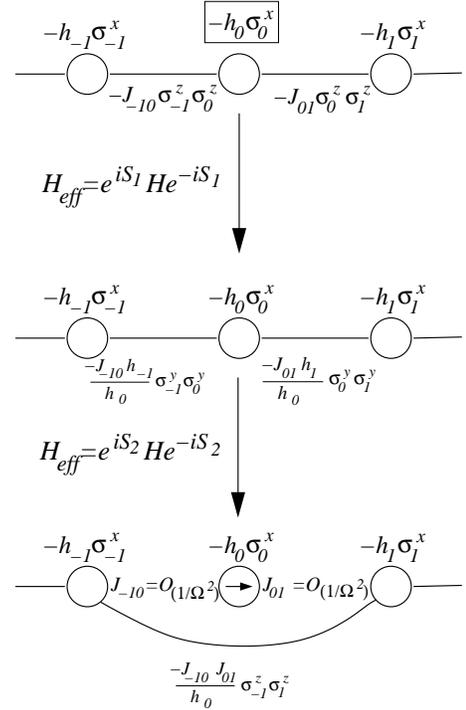}
\caption{Site decimation. Spin 0 is almost frozen in the $x$-direction due to the strong magnetic field $h_0$. Quantum 
fluctuations create a second 
nearest neighbor effective interaction between sites -1 and 1. This interaction is weaker than any of $J_{-10},\,J_{01},\,h_0$. 
\label{hdecfig}}
\end{figure}

We now apply the transformations (\ref{Stransform}) to successively
reduce the maximum energy scale of the random quantum Ising Hamiltonian, thereby obtaining a series of low energy effective Hamiltonians of the system.
We begin by choosing the largest energy coefficient in the Hamiltonian
(\ref{eq:tfisc}) and denote it $\Ho$ (with the corresponding coupling
the initial energy scale $\Omega_I)$: for example,  $\Ho=-h_{1}\sx{1}$
(see Fig. \ref{hdecfig}). Let $V$ designate the part of the
Hamiltonian that we would like to eliminate. For the above example we
would like to eliminate the parts connecting site $1$ to the rest of
the chain: 
\be
V=-J_{01}\sz{0}\sz{1}-J_{12}\sz{1}\sz{2}.
\ee
In addition to these two parts, the Hamiltonian also contains the parts involving the remainder of the chain,
\be
\Hl=\ldots-J_{-10}\sz{-1}\sz{0}-h_{0}\sx{0}-h_{2}\sx{2}-J_{23}\sz{2}\sz{3}+\ldots
\ee

In order to eliminate the first order couplings to spin 0, $S$ must satisfy equation (\ref{eq:comm}); thus we first choose
\be
S_{a}=-\frac{J_{01}}{2h_{1}}\sz{0}\sy{1}-\frac{J_{12}}{2h_{1}}\sy{1}\sz{2}   \label{S1hdec}
\ee
which yields the following terms in the effective Hamiltonian (Eq. \ref{eq:heff}):
\be
\ba{c}
\H_{eff}=\vspace{2mm}\\
\ldots-\frac{J_{01}J_{12}}{h_{1}}\sz{0}\sx{1}\sz{2}
	-\frac{h_{0}J_{01}}{h_{1}}\sy{0}\sy{1}-\frac{h_{2}J_{12}}{h_{1}}\sy{1}\sy{2}+\ldots
\ea
\ee
Note that site  1 is still coupled to adjacent sites by a second order interaction. We would like to restore the Hamiltonian to its original form; thus we need to eliminate the new type of interaction. To get rid of it, we perform another transformation using 
\be
S_{b}=-\frac{h_{0}J_{01}}{2h_{1}^{2}}\sy{0}\sz{1}-\frac{h_{2}J_{12}}{2h_{1}^{2}}\sz{1}\sy{2}. \label{S2hdec}
\ee
The effective Hamiltonian now includes
\be
\begin{array}{c}
\H_{eff}=\ldots-J_{-10}\sz{-1}\sz{0}-h_{0}\sx{0}\vspace{2mm}\\
-\tilde{J}_{02}\sz{0}\sx{1}\sz{2}-h_{2}\sx{2}-J_{23}\sz{2}\sz{3}+\ldots,\vspace{2mm}\\
-h_{1}\sx{1},
\ea
\label{HHH1}
\ee
from which we see that in the low-energy subspace of $\H_0$, the effective exchange between spins 0 and 2 is
given by
\be
\tilde{J}_{02}=\frac{J_{01}J_{12}}{h_{1}}.
\label{je1}
\ee
Since $h_1$ is the strongest
coupling energy in the chain, the resulting effective bond obeys 
\be
\tilde{J}_{02}\ll h_0,\,J_{01},\,J_{12},
\ee
where the sharpness of the inequality is because we assume strong
randomness. We can now partially diagonalize $\H_{eff}$ by writing
\be
\ba{c}
\pni=|\rightarrow>_{1}|\tilde{G}^{(1)}>\vspace{2mm}\\
\pgs=e^{-iS_a}e^{-iS_b}\pni ,
\ea
\ee
where $\sx{1}|\rightarrow_{1}>=|\rightarrow_{1}>$ and
$|\tilde{G}^{(1)}>$ involves only the spins other than 1.  We are left
with a renormalized spin-chain with the spin at site 1 eliminated, and
with an effective interaction $\tilde{J}_{02}\sz{0}\sz{2}$ between
spin 0 and spin 2.
[Note that we could also keep the high energy sector that involves
  $|\leftarrow_1>$; the effective Hamiltonian and state of the rest of
  the chain would differ from those of the low energy sector because
  of the presence of $\sx{(1)}$ in $\H_{eff}$ of Eq. \ref{HHH1}.]

% ----------------------J-dec results%

The analog of the above results for the case where an exchange interaction, e.g.  $\Ho=-J_{12}\sz{1}\sz{2}$, is eliminated is (Fig. \ref{Jdecfig})
\be 
\begin{array}{c}

\Ho=-J_{12}\sz{1}\sz{2}\vspace{2mm}\\
V=-h_{1}\sx{1}-h_{2}\sx{2}\vspace{2mm}\\
S_a=\frac{h_{1}}{2J_{12}}\sy{1}\sz{2}+\frac{h_{2}}{2J_{12}}\sz{1}\sy{2}\vspace{2mm}\\   \label{jdec}
S_b=-\frac{h_{0}J_{01}}{2J_{12}^{2}}\sy{1}\sz{0}-\frac{h_{3}J_{23}}{2J_{12}^{2}}\sz{3}\sy{2}.
\ea
\ee
This could be obtained by using the duality described in the
introduction (Sec. \ref{intdual}) and in Ref. \onlinecite{DSF95}, or by direct computation.  
The ground state of $\H_0=-J_{12}\sz{1}\sz{2}$ is doubly degenerate 
with spins 1 and 2 either in the state $|\uparrow_{(12)}>=|\uparrow_{1}>|\uparrow_{2}>$ 
or in the state
$|\downarrow_{(12)}>=|\downarrow_{1}>|\downarrow_{2}>$. Therefore in
the ground state of $\H_0=\sz{1}\sz{2}$ spin 1 and 2 form a
ferromagnetic cluster, which we denote as $(12)$. We can define cluster  operators, $\sz{(12)}$ and 
$\sx{(12)}$, that operate on the spin cluster $(12)$ in the following way:
\be
\ba{rcl}
&\sz{1}\Rightarrow\sz{(12)}& \nonumber \vspace{2mm}\\ &\sz{2}\Rightarrow\sz{(12)}& \nonumber \vspace{2mm}\\ & -\sy{1}\sy{2}\Rightarrow \sx{(12)} 
\ea
\ee
in terms of which
\be
\begin{array}{c}
\H_{eff}-\Ho=\vspace{2mm}\\
\ldots-h_{0}\sx{0}-J_{0(12)}\sz{0}\sz{(12)}-\tilde{h}_{(12)}\sx{(12)}\vspace{2mm}\\
-J_{(12)3}\sz{(12)}\sz{3}-h_{3}\sx{3}-\ldots,
\end{array}
\label{he1}
\ee
with
\be
\tilde{h}_{(12)}=\frac{h_{1}h_{2}}{J_{12}}
\ee
being the effective transverse field on the new cluster $(12)$ that has
replaced the pair of spins 1 and 2 that now only appear separately in
the high energy term in $\Ho$.  Again, since $J_{12}$ is the strongest
energy, and strong randomness is assumed, the effective transverse
field obeys:
\be
\tilde{h}_{(12)}\ll h_1,\,h_2,\,J_{12}.
\ee

\begin{figure}
\includegraphics[width=7.5cm]{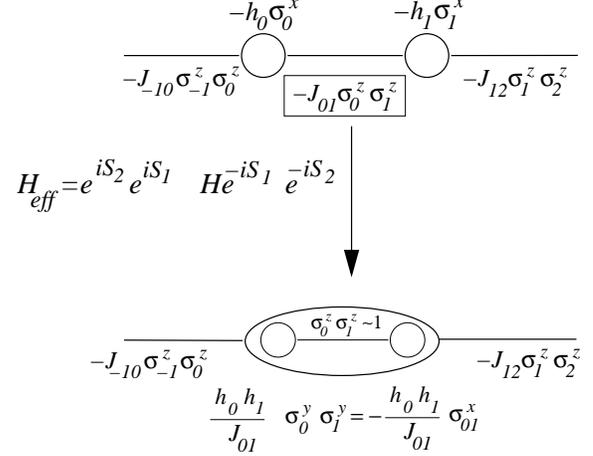}
\caption{Bond decimation. Sites 0 and 1 are frozen into one cluster by the strong Ising interaction, $J_{01}$. 
Quantum fluctuations produce an effective magnetic field,
$\tilde{h}_{01}=\frac{h_0 h_1}{J_{01}}$, which flips the composite
spin cluster. This field is weaker than any of $h_0,\,h_1,\,J_{01}$.
\label{Jdecfig}} 
\end{figure}

In both the decimation cases we regain the initial form of the
Hamiltonian, but with one less spin. As shown in
Refs. \onlinecite{DSF94, DSF95}, with even stronger randomness. The
increase in the randomness with each step justifies the iterative
application of the real-space RG as described in this section. In each
step we eliminate a high-energy subspace of the Hilbert space of the chain, which
is gapped by $2\Omega$ from the remaining subspace. The range of
excitations in the remaining subspace is much smaller than
$\Omega$. The iterative application of the decimation procedure
outlined here amounts to separating the Hilbert space of the chain
into a hierarchy of sequentially decreasing energy subspaces. If the
coupling distributions expand without bounds during the flow, this method is
asymptotically exact.

After applying this set of transformation rules $L$ (the original
chain length) times, we are left with a single spin cluster that
carries the moments of some fraction of the set of original
spins. The ground state of the chain is then given by the state in
which this cluster points in the $x$ direction due to
the transverse field. In the same way we can also access the various
excitations of the quantum Ising chain by keeping high energy subspaces
in the decimation process.

At the end of the decimation process, the full effective Hamiltonian is given by
\begin{widetext}
\be
\H_{eff}=e^{iS^{(L)}}e^{iS^{(L-1)}}\ldots e^{iS^{(2)}}e^{iS^{(1)}}\H e^{-iS^{(1)}} e^{-iS^{(2)}}\ldots e^{-iS^{(L-1)}} e^{-iS^{(L)}},
\ee
\end{widetext}
with $S^{(j)}$ representing the transformation of the j'th stage of the renormalization. 
At the final stage the free ground-state wave-function, $\pni$, is related to the ground state of the original problem by
\be
\pgs=e^{-iS^{(1)}}e^{-iS^{(2)}}\ldots e^{-iS^{(L-1)}} e^{-iS^{(L)}}\pni.
\ee
Note that the later transformations are the first to operate on
$\pni$. In the end of the decimation process, $\pni$ is a product of
cluster wave functions, where each cluster points in the direction of
the transverse field. For example, $\pni$ for a chain of 4 spins as in
Fig. \ref{fig3-5} is given by
\be
\ba{c}
\pni=|\rightarrow_{(023)}>|\rightarrow_1>\vspace{2mm}\\
=\l(|\uparrow_0>|\uparrow_2>|\uparrow_3>+|\downarrow_0>|\downarrow_2>|\downarrow_3>\r)\l(|\uparrow_1>+\downarrow_1>\r).
\ea
\label{pniex}
\ee

\begin{figure}
\includegraphics[width=7cm]{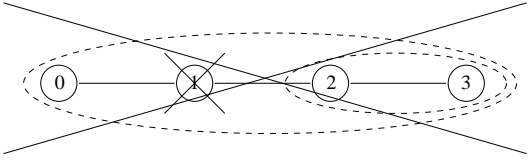}
\caption{Example of hierarchical decimation of a chain with four
  sites. First, site $1$ is decimated. At a lower energy scale, sites
  2 and 3 form a ferromagnetic cluster, which we denote (23). Cluster (23) then
  forms a cluster with site 0. The last process is a decimation of the
  cluster (023). The gound state wvae function of this chain is
  constructed perturbatively from the hierarchical wavefunction
  $\pni$, which is given in Eq. (\ref{pniex}). \label{fig3-5}}
\end{figure}

% --------------------------------------effective operators%

\subsection{Evolution of Effective Operators}

The quantities we are interested in can be written in terms of $\pni$ and the set of unitary transformations used in the RG process. For example, let us consider
\be
<\sx{0}\sx{L}>=<G|\sx{0}\sx{L}\pgs,
\ee
with $\pgs$ known in terms of $\pni$. We can write
\be
\begin{array}{c}
<G|\sx{0}\sx{L}\pgs=\vspace{2mm}\\
<H|e^{iS^{(L)}}\ldots e^{iS^{(2)}}e^{iS^{(1)}}\sx{0}\sx{L}e^{-iS^{(1)}}e^{-iS^{(2)}}\ldots e^{-iS^{(L)}}\pni\vspace{2mm}\\
=<H|e^{iS^{(L)}}\ldots e^{iS^{(2)}}e^{iS^{(1)}}\sx{0}e^{-iS^{(1)}}e^{-iS^{(2)}}\ldots e^{-iS^{(L)}}\vspace{2mm}\\
e^{iS^{(L)}}\ldots e^{iS^{(2)}}e^{iS^{(1)}}\sx{L}e^{-iS^{(1)}}e^{-iS^{(2)}}\ldots e^{-iS^{(L)}}\pni\vspace{2mm}\\
=<H|\tilde{\sx{0}}\tilde{\sx{L}}\pni
\end{array}
\ee

Thus we see that calculating expectation values of an operator $A$ with respect to the ground state $\pgs$ is equivalent to calculating the expectation value of the effective operator $\tilde{A}$ with respect to $\pni$:
\be
\begin{array}{c}
<G|A\pgs=<H|\tilde{A}\pni,\vspace{2mm}\\
\tilde{A}= e^{iS^{(L)}}\ldots e^{iS^{(2)}}e^{iS^{(1)}}A~e^{-iS^{(1)}}e^{-iS^{(2)}}\ldots e^{-iS^{(L)}}.
\end{array}
\ee

%------------------end of formalism,    +++++++++++++++     begin sx-sx results%

\section{Transverse Field Correlations \label{sx-sx}}

%-------------------------------RG of spin end operatoers  --------------5%

\subsection{Renormalization of End Spin Operators}

In order to obtain the transverse field part of the end-to-end energy
correlations, $<h_0\sx{0}h_L\sx{L}>$, we must consider the effects of
the two types of renormalization steps (decimation of a site or a
bond) on the end operators $\sx{0}$ and  $\sx{L}$. In this section we
show that the decimation of the first bond  makes the operator $\sx{0}$ evolve to the
effective operator
$\tilde{\sx{0}}=\frac{J_{01}}{h_0}\sx{(01)}$. We also show that the Decimation of the end site
yields $\tilde{\sx{0}}=\frac{J_{01}^2h_1}{h_0^3}\sx{1}$, which is
third order in the
off-diagonal terms that couple high and low energies in the
Hamiltonian. In this derivation we neglect all the subleading
contributions to the flow of the edge operatros; we show in Appendix
\ref{appA} that this is indeed justified.

%-------------------chain body rg%

\paragraph{Decimations away from the ends.}

A renormalization step that does not involve the end spin will
generally leave the end operators unchanged, since the generator of the
unitary transformation of this RG step, $S$, commutes with $\sx{0}$.
The exception to this is when the decimation involves spins {\it
  adjacent} to the end, for which $[S,\sx{0}]\neq 0$; however, it can
be shown that to all orders in perturbation theory, there is no
contribution to the dominant parts of the correlation function from
the resulting corrections to $\sx{0}$.

%-----------------------------------rg of end%

%----------------------------------------------J-dec%

\paragraph{Decimation of an end bond.}
\begin{figure}
\includegraphics[width=6cm]{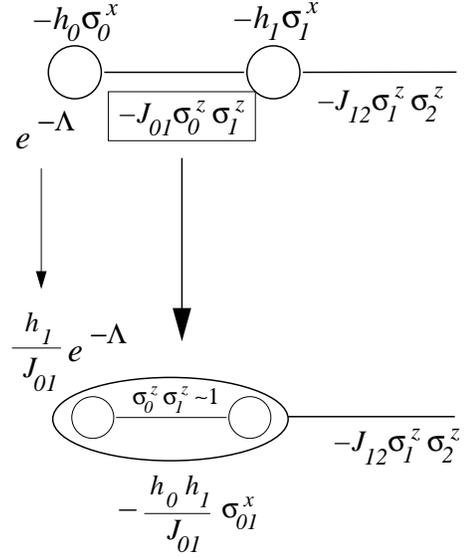}
\caption{Evolution of $\sigma^x_0$ in a bond decimation. When the end site forms a cluster with its neighbor, the 
operator $\sigma^x_0$ gets renormalized and gains a factor of $h_1/J_{01}$.  \label{endJfig}}
\end{figure} 

If $J_{01}$ is decimated, sites 0 and 1 will form a cluster $(01)$ (see Fig. \ref{endJfig}).
The dominant contribution to the correlation function comes from the
effective operator $\sx{(01)}$.  This contribution is obtained from the first order transformation, $S_a$ in Eq. (\ref{jdec}):
\be
\ba{l}
\tilde{\sx{0}}=e^{iS_a}\sx{0}~e^{-iS_a}=\vspace{2mm}\\
\hspace{10mm}\sx{0}+i[\frac{h_{0}}{2J_{01}}\sy{0}\sz{1},\sx{0}]+i[\frac{h_{1}}{2J_{01}}\sy{1}\sz{0},\sx{0}]=\vspace{2mm}\\ \label{xJ}
\hspace{10mm}\sx{0}+\frac{h_{0}}{J_{01}}\sz{0}\sz{1}-\frac{h_{1}}{J_{01}}\overbrace{\sy{1}\sy{0}}^{-\sx{(01)}}.
\ea
\ee

The first two terms in (\ref{xJ}) will not evolve under the continuing renormalization. Their expectation values are
\be
\ba{c}
<H|\sx{0}\pni=0,\vspace{2mm}\\
<H|\sz{0}\sz{1}\pni=1.
\ea
\ee

The only piece of (\ref{xJ}) relevant to us is the third
piece. As indicated in Eq. (\ref{xJ}), when the operator
$-\sy{1}\sy{0}$ is restricted to the low energy subspace in which sites
0 and 1 form a cluster, it is equivalent to the cluster operator
$\sx{(01)}$. Therefore when the end spin forms a
cluster with its neighbor via a bond decimation, the flow of the
transverse spin is given by
\be
\sx{0}\Rightarrow\frac{h_{1}}{J_{01}}\sx{(01)}. \label{relxJ}
\ee

%----------------------------------   h-dec%

\paragraph{End site decimation.}

If $h_{0}$ is the strongest interaction in the chain, site 0 will be
decimated. Applying the first order transformation $S_a$  from
Eq. (\ref{S1hdec}) to $\sx{0}$ yields
\be
\ba{c}
S_a=-\frac{J_{01}}{2h_{0}}\sy{0}\sz{1}\vspace{2mm}\\
\tilde{\sx{0}}=\sx{0}+i[\frac{J_{01}}{2h_{0}}\sy{0}\sz{1},\sx{0}]=\sx{0}+\frac{J_{01}}{h_{0}}\sz{0}\sz{1}.
\ea
\ee
\begin{figure}
\vspace{3mm}
\includegraphics[width=6cm]{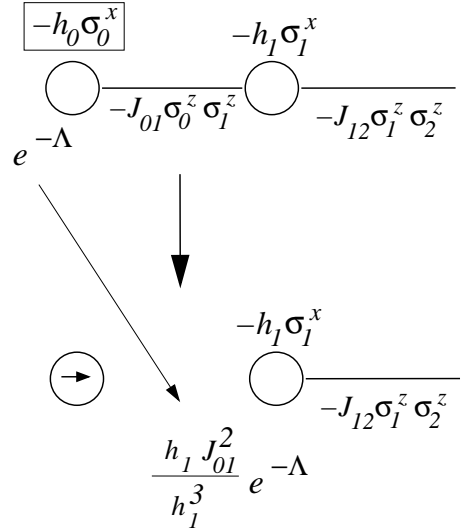}
\caption{Evolution of $\sigma^x_0$ in a site decimation. Although $\sigma_0^x$ obtains an expectation value, its fluctuations are 
still influenced by the state of site 1. This is reflected in  the renormalization of $\sigma_0^x$ as 
$\frac{J_{01}^2}{h_0^2}\sigma_1^x$ which  gives rise  factor of $\frac{J_{01}^2h_0}{h_1^3}$ 
 in the correlation functions of $\sx{0}$.
\label{endhfig}} 
\end{figure} 
Applying the second order transformation yields 
\be
\ba{c}
S_b=-\frac{h_{1}J_{01}}{2h_{0}^{2}}\sz{0}\sy{1},\vspace{2mm}\\\label{hx}
\tilde{\sx{0}}=\sx{0}-i[\frac{h_{1}J_{01}}{2h_{0}^{2}}\sz{0}\sy{1},\sx{0}+\frac{J_{01}}{h_{0}}\sz{0}\sz{1}]\vspace{2mm}\\
=\sx{0}+\frac{h_{1}J_{01}}{h_{0}^{2}}\sy{0}\sy{1}+\frac{h_{1}J_{01}^{2}}{h_{0}^{3}}\sx{2}.
\ea
\ee
Once again, the first two terms in Eq. (\ref{hx}) do not contribute to
truncated correlation functions since $<\sx{0}>=1$, $<\sy{0}>=0$. The
flow of $\sx{0}$ in this case is 
\be
\sx{0}\Rightarrow\frac{h_{1}J_{01}^{2}}{h_{0}^{3}}\sx{1}. \label{relxh}
\ee

%- ----------------------    keeping track -  OMEGA%

\subsection{Evolution of the Correlation Function \label{evsec}}

As the renormalization proceeds, the effective operators $\tilde\sx{0
}$ and $\tilde\sx{L}$ accrue multiplicative  factors that will
eventually combine to form the end-to-end correlation function. The
evolution of these prefactors is obtained from Eqs. (\ref{relxJ}) and
(\ref{relxh}) by the method outlined in Ref. \onlinecite{DSF98}. 

At log-energy scale $\Gamma$, we write the effective operator of the end spin as  \be
\sx{0}\Rightarrow\tilde\sx{0}(\Gamma)e^{-\Lambda(\Gamma)},
\ee
where $\sx{0}(\Gamma)$ operates on the first spin cluster of the renormalized chain at the scale $\Gamma$. 
Using the logarithmic variables
$\beta=\ln\frac{\Omega}{h},~\zeta=\ln\frac{\Omega}{J}$, we can rewrite
Eqs. (\ref{relxJ}, \ref{relxh}) and the results described in
Figs. (\ref{endJfig}) and (\ref{endhfig}) in terms of $\Lambda$: 
\be
\ba{ll}
\mbox{bond-decimation:    }   &  \Lambda\Rightarrow\Lambda+\beta_{1},\vspace{2mm}\\\label{lambdaev}
\mbox{spin-decimation:    }   &  \Lambda\Rightarrow\Lambda+{\beta_{1}+2\zeta_{01}}.
\ea
\ee
The quantities $\Lambda$, $\beta_{0}$, and $l_{0}^{c}$ (length of the
cluster containing the end spin) are correlated at all stages of the
renormalization. Therefore we must keep track of their joint
distribution, which we define as
\be
\rm{Prob}[dl_{0}^{c},d\beta_{0},d\Lambda]=\omega(\beta_{0},l_{0}^{c},\Lambda|\Gamma)d\beta_{0}dl_{0}^{c}d\Lambda.
\ee
Using the results of ref. \onlinecite{DSF95} and Eqs. (\ref{jdec}) and
(\ref{lambdaev}), we can write the evolution equation for $\omega(\beta,l,\Lambda|\Gamma)$:
\begin{widetext}
\be
\ba{l}
\frac{d\omega{(\beta,l,\Lambda|\Gamma)}}{d\Gamma}=\frac{\partial\omega{(\beta,l,\Lambda|\Gamma)}}{\partial \beta}\vspace{2mm}\\
+\int\omega{(\beta_{0}=0,l_{c}^{0},\Lambda'|\Gamma)}P{(\zeta_{0},l^{b}_{0})}R{(\beta_{1},l^{c}_{1})}\delta{(\Lambda-\Lambda'-2\zeta_{0}-\beta_{1})}\delta{(l-l_{0}^{c}-l_{0}^{b}-l_{1}^{c})}\delta{(\beta-\beta_{1})}dl_{0}^{c}dl_{0}^{b}dl_{1}^{c}d\Lambda' d\beta_{1}d\zeta_0\vspace{2mm}\\
+\int\omega{(\beta_{0},l_{0}^{c},\Lambda'|\Gamma)}R{(\beta_{1},l_{1}^{c})}P{(\zeta_{0}=0,l_{0}^{b})}\delta{(\Lambda-\Lambda'-\beta_{1})}\delta{(l-l_{0}^{c}-l_{0}^{b}-l_{1}^{c})}\delta{(\beta-\beta_{1}-\beta_{2})}dl_{0}^{c}dl_{0}^{b}dl_{1}^{c}d\Lambda' d\beta_{0}d\beta_{1}\vspace{2mm}\\
-\int P{(0,l')}dl'\omega{(\beta,l,\Lambda|\Gamma)}.
\ea
\ee

in terms of the distributions $P$ and $R$ of the log-couplings and
lengths of the bonds and spin-clusters at scale $\Gamma$. Employing
the notation for convolutions introduced in Ref. \onlinecite{DSF95}, 
$f{(x_1)}\stackrel{x}{\otimes}g{(x_2)}=\int
f{(x_1)}g{(x_2)}\delta{(x-x_1-x_2)}dx_1dx_2$, we can write the
above equation in a more compact way:
\be
\ba{c}
\frac{d\omega{(\beta,l,\Lambda|\Gamma)}}{d\Gamma}=\frac{\partial\omega{(\beta,l,\Lambda|\Gamma)}}{\partial \beta}\vspace{2mm}\\
+\int\omega{(\beta_{0}=0,l_{c}^{0},\Lambda'|\Gamma)}\stackrel{l}{\otimes}P{(\zeta_{0},l^{b}_{0})}\stackrel{l}{\otimes}R{(\beta,l^{c}_{1})}\delta{(\Lambda-\Lambda'-2\zeta_{0}-\beta)}d\Lambda' d\zeta_{0}\vspace{2mm}\\
+\int\omega{(\beta_{0},l_{0}^{c},\Lambda'|\Gamma)}\stackrel{\beta,l}{\otimes}R{(\beta_{1},l_{1}^{c})}\stackrel{l}{\otimes}P{(\zeta_0,l_{0}^{b})}\delta{(\Lambda-\Lambda'-\beta_{1})}d\Lambda'\vspace{2mm}\\
-\int P{(0,l')}dl'\omega{(\beta,l,\Lambda|\Gamma)}.
\ea
\ee

\end{widetext}

By Laplace transforming $\omega{(\beta,l,\Lambda)}$ with respect to
both $l$ and $\Lambda$, 
\be
\omega{(\beta,y,\lambda)}=\int dl\int d\Lambda e^{-ly} e^{-\lambda\Lambda} \omega{(\beta,l,\Lambda)}, 
\ee
we obtain
\be
\ba{c}
\frac{d\omega{(\beta,y,\lambda|\Gamma)}}{d\Gamma}=\frac{\partial\omega{(\beta,y,\lambda)}}{\partial \beta}\vspace{2mm}\\
+\omega{(\beta_{0}=0,y,\lambda|\Gamma)}\int e^{-2\zeta\lambda}P{(\zeta,y)}d\zeta
R{(\beta,y)}\vspace{2mm}\\
+\int\omega{(\beta_{0},y,\lambda)}\stackrel{\beta}{\otimes}R{(\beta_{1},y)}P{(\zeta_{0}=0,y)}\vspace{2mm}\\
-P{(0,y=0)}\omega{(\beta,y,\lambda)}.
\ea
\ee
In Ref. \onlinecite{DSF95} the scaling limits of the functions
$R{(\beta,\,y)},\,P{(\beta,\,y)}$ are derived:
\be
\ba{c}
P{(\zeta,y|\Gamma)}=\Upsilon{(y,\Gamma)} e^{-\zeta
  u{(y,\Gamma)}},\vspace{2mm}\\
R{(\beta,y|\Gamma)}=T{(y,\Gamma)} e^{-\beta \tau{(y,\Gamma)}},\vspace{2mm}\\
T{(y,\Gamma)}=\frac{\Delta{(y)}}{\sinh\Delta{(y)}\Gamma}
e^{\delta\Gamma},\vspace{2mm}\\
\Upsilon{(y,\Gamma)}=\frac{\Delta{(y)}}{\sinh\Delta{(y)}\Gamma}
e^{-\delta\Gamma},\vspace{2mm}\\
\tau{(y,\Gamma)}=\delta+\Delta{(y)}\coth\l(\Delta{(y)}\Gamma\r),\vspace{2mm}\\
u{(y,\Gamma)}=-\delta+\Delta{(y)}\coth\l(\Delta{(y)}\Gamma\r),\vspace{2mm}\\
\Delta{(y)}=\sqrt{y+\delta^2}.
\ea
\ee
Using these results and the corresponding notations,
 we get
\be
\ba{c}
\frac{d\omega{(\beta,y,\lambda)}}{d\Gamma}=\frac{\partial\omega{(\beta,y,\lambda)}}{\partial \beta}
+\omega{(0,\lambda,y)}\frac{T{(y,\Gamma)}\Upsilon{(y,\Gamma)}}{2\zeta+u}\vspace{2mm}\\
+\Upsilon{(y,\Gamma)}\int\omega{(\beta_{0},y,\lambda)}e^{-\tau\beta_{1}-\lambda\beta_{1}}T{(y,\Gamma)}\vspace{2mm}\\
\cdot\delta{(\beta-\beta_{0}-\beta_{1})}d\beta_{0}d\beta_{1}\vspace{2mm}\\
-\Upsilon{(0,\Gamma)}\omega{(\beta,y,\lambda)} \label{omegae}.
\ea
\ee
To solve this we write $\omega{(\beta,y,\lambda)}$ in the following form:
\be
\omega{(\beta,y,\lambda)}=W{(y,\lambda)}e^{-\beta\lambda-\beta\tau},
\ee
under which (\ref{omegae}) becomes
\be
\frac{dW}{d\Gamma}=-(\tau+\lambda)W+\frac{\Upsilon{(y,\Gamma)}T{(y,\Gamma)}}{2\lambda+u}W-\Upsilon{(0,\Gamma)}W .   \label{omdif}
\ee
Using the definitions of the functions $\Upsilon,\,T,\,u$, and $\tau$
, we can integrate Eq. (\ref{omdif}) and find
\be
\ba{c}
W=W_{0}e^{(\lambda+\delta)(\Gamma-\Gamma_I)}\vspace{2mm}\\
\cdot\frac{\sinh(\Delta{(y)}\Gamma_{I})}{\sinh(\Delta{(y)}\Gamma)}\left(\frac{2\lambda+u{(y,\Gamma_{I})}}{2\lambda+u{(y,\Gamma)}}\right)\frac{\tau{(0,\Gamma)}}{\tau{(0,\Gamma_{I})}}.  
\label{Omega}
\ea
\ee
The value of $W_0$ can be found from the normalization condition
on the distribution $\omega{(\beta,l,\Lambda)}$.  To be a properly
normalized,   $\omega{(\beta,l,\Lambda)}$ has to obey
\be
\int\limits_{0}^{\infty}d\beta\omega{(\beta,y=0,\lambda=0)}=1.
\ee 
This implies $W_{0}=\tau{(0,\Gamma_{I})}$. 

Before  proceeding, we should check what initial conditions this distribution satisfies. Setting $\Gamma\Rightarrow\Gamma_I$ and $y\Rightarrow 0$  we see:
\be
\omega{(\beta,0,\lambda|\Gamma_I)}=\tau{(0,\Gamma_{I})}e^{-\tau{(y,\Gamma_{I})}\beta}e^{-\lambda\beta}
\ee
The inverse Laplace Transform in $\lambda$ yields
\be
\omega{(\beta,0,\Lambda|\Gamma_I)}=\tau{(0,\Gamma_{I})}e^{-\tau{(0,\Gamma_{I})}\beta}\delta{(\Lambda-\beta)}.\label{om1}
\ee
The delta function in Eq. (\ref{om1}) shows that the initial value of
the correlation variable is the same as the transverse field, as it
should be for this part of the energy - energy correlations.

%---------------------------------------J(lambda,L) part

\subsection{Last Decimation Step}

After carrying out the decimation process $L-2$ times and following
the flow of the edge energy operators, we end up with two
clusters. The remaining clusters correspond to the left and right
edges of the chain, and each has a distribution
$\omega{(\beta,l,\Lambda|\Gamma)}$ associated with it. In the next
decimation step one of these clusters is decimated, and a single
cluster forms. At this stage we can compute the distribution of the
energy correlations from the flow of edge operators and the remaining
couplings. Note that
the $\Gamma$ at which a single cluster forms is logarithm of the gap
between the first and second excited states (as was also noted in
Ref. \onlinecite{igloi}). 

The computation of the truncated correlation function
is as follows:
\be
\ba{c}
C^{xx}_L=<G|h_0\sx{0}h_L\sx{L}\pgs\vspace{2mm}\\
-<G|h_0\sx{0}\pgs <G|h_L\sx{L}\pgs\vspace{2mm}\\   \label{finalxx}
\hspace{10mm}=<H|e^{-\Lambda_{\ell}}\sx{\tilde{\ell}}\sx{\tilde{r}}e^{-\Lambda_{r}}\pni\vspace{2mm}\\
-<H|e^{-\Lambda_{\ell}}\sx{\tilde{\ell}}\pni <H|\sx{\tilde{r}}e^{-\Lambda_{r}}\pni ,
\ea
\ee
where $\Lambda_{\ell}$ and $\Lambda_{r}$ are the correlation factors
picked up in the RG process, Eq. $(\ref{lambdaev})$, for the left end
and right end transverse spins respectively, and we have labeled the
last remaining spin clusters $\ell$ and $r$.
The correlation function can also be written as a sum over excited states:
\be
\ba{c}
C_L^{xx}=
<h_0\sx{0}h_L\sx{L}>-<h_0\sx{0}><h_L\sx{L}>\vspace{5mm}\vspace{2mm}\\
=<G|(h_0\sx{0}-<h_0\sx{0}>)(h_L\sx{L}-<h_L\sx{L}>)\pgs\vspace{5mm}\vspace{2mm}\\
=\sum\limits_{\psi\neq G}<G|h_0\sx{0}|\psi><\psi|h_L\sx{L}\pgs\vspace{5mm}\vspace{2mm}\\
=e^{-\Lambda_{\ell}-\Lambda_r}\sum\limits_{\psi\neq G}<H|\sx{\tilde{0}}|\psi><\psi|\sx{\tilde{1}}\pni ,
\ea \label{likesum}
\ee
where the sum over $\psi$ runs over all states except the ground state.

At the penultimate decimation step two processes are possible:

\paragraph{Bond Decimation.} The two remaining clusters become a
single cluster, and  the only remaining coupling is the transverse
field, which makes the combined cluster point in the x direction:
\be
|H^{(L)}>=\frac{1}{\sqrt{2}}(|\uparrow_{\tilde{\ell}}>|\uparrow_{\tilde{r}}>+|\downarrow_{\tilde{\ell}}>|\downarrow_{\tilde{r}}>),
\ee
where $\tilde{\ell}$ and $\tilde{r}$ represent the  end clusters at this final stage of the RG.
This yields (\ref{finalxx}) 
\be
C_L^{xx}=e^{-\Lambda_{\ell}-\Lambda_{r}}=e^{-\Lambda}; 
\ee
hence
\be
\Lambda=\Lambda_{r}+\Lambda_{\ell},  \label{psa}
\ee
where $\Lambda$ is the desired log of the energy-correlations. 

\paragraph {Site Decimation.} In the case of a site decimation it is unimportant which of the last
surviving clusters gets decimated. Let us assume that it is the left
cluster, $\tilde{\ell}$. This involves the unitary transformation (dropping the
tildes) $S_a=-\frac{J_{\ell
    r}}{\tilde{h}_{\ell}}\sy{\tilde{\ell}}\sz{\tilde{r}}$, and makes
the ground state $\pni$ be
\be
\H^{(L)}>=|\rightarrow_{\ell}>|\rightarrow_{r}>=\frac{1}{2}(|\uparrow_{\ell}>+|\downarrow_{\ell}>)(|\uparrow_{r}>+|\downarrow_{r}>).
\ee
By using the sum form in Eq. (\ref{likesum}) and applying $S_a$ we get
\begin{widetext}
\be
\ba{c}
C_L^{xx}e^{\Lambda_l+\Lambda_r}\vspace{2mm}\approx\sum\limits_{\psi\neq G}<H|e^{iS_a}\sx{\ell}e^{-iS_a}|\psi><\psi|e^{iS_a}\sx{r}e^{-iS_a}\pni\vspace{2mm}\\
\approx\sum\limits_{\psi\neq G}<H|-\frac{J_{\ell r}}{h_\ell}\sz{\ell}\sz{r}|\psi><\psi|\frac{J_{\ell r}}{h_\ell}\sy{\ell}\sy{r}\pni=\frac{J_{\ell r}^2}{h_\ell^2}.
\ea
\ee
This yields
\be
\Lambda=\Lambda_{r}+\Lambda_{\ell}+2\zeta_{\tilde{\ell}\tilde{r}}.  \label{psb}
\ee

The analog of Eq. (25) in Ref. \onlinecite{DSF98} for the measures defined here is
\be
\ba{l}
d{\rm Prob}(\Lambda_{\ell},l_{\ell},\beta_{\ell},l^{b}_{1},\zeta_{1},l^{c}_{1},\beta_{1},l^{b}_{2},\zeta_{2},\ldots,l_{r},\beta_{r},\Lambda_{r}|L,\Gamma)=\vspace{2mm}\\
a_{\Gamma}\omega{(\beta_{\ell},\Lambda_{\ell},l_{\ell})}P{(\zeta_{1},l^{b}_{1})}R{(\beta_{1},l^{c}_{1})}P{(\zeta_{2},l^{b}_{2})}\ldots\omega{(\beta_{r},\Lambda_{r},l_{r})}\delta{(l_{\ell}+l^{b}_{1}+l^{c}_{1}+l^{b}_{1}+\ldots+l_{r}-L)}d\{\beta_i\}d\{\zeta_i\}d\Lambda_rd\Lambda_{\ell}d\{l_i\},  \label{dprob}
\ea
\ee
with:
\be
\frac{1}{a_{\Gamma}}\frac{da_{\Gamma}}{d\Gamma}=\int\limits_{0}^{\infty}(P{(0,l)}+R{(0,l)})dl.
\ee

Let us now define the function $J{(\Lambda,\Gamma|L)}$ as
\be
d{\rm Prob}[\mbox{chain of length $L$ becomes a single cluster at $\Gamma$ with $\log(C_{L})=\Lambda$}]=J{(\Lambda,\Gamma|L)}d\Gamma d\Lambda.
\ee
The probability distribution of $\log(C_L)$   is then given by
 \be 
f{(\Lambda,L)}=\int\limits_{0}^{\infty}J{(\Lambda,\Gamma|L)}d\Gamma.
\ee

The function  $J{(\Lambda,\Gamma|L)}$ has two contributions, the
first contribution cames from the case of the penultimate decimation
being a bond decimation (Eq. \ref{psa}). The second contribution
comes from the case of a site-decimation (Eq. \ref{psb}). The
combination of the two contribution yields:
\be
\ba{c} \label{Jps}
J{(\Lambda,\Gamma|L)}=\vspace{2mm}\\
a_{\Gamma}\int P{(0,l^{b}_{\ell r})}\omega{(\beta_{\ell},\Lambda_{\ell},l_{\ell})}\omega{(\beta_{r},\Lambda_{r},l_{r})}\delta{(l_{\ell}+l^{b}_{\ell r}+l_{r}-L)}\delta{(\Lambda-\Lambda_{\ell}-\Lambda_{r})}
d\Lambda_{\ell}d\Lambda_{r}dl_{\ell}dl^{b}_{\ell r}dl_{r}d\beta_{r}d\beta_{\ell}\vspace{2mm}\\
+2a_{\Gamma}\int P{(\zeta_{r},l^{b}_{\ell r})}\omega{(0,\Lambda_{\ell},l_{\ell})}\omega{(\beta_{r},\Lambda_{r},l_{r})}\delta{(l_{\ell}+l^{b}_{\ell r}+l_{r}-L)}\delta{(\Lambda-\Lambda_{\ell}-\Lambda_{r}-2\zeta_{r})}
d\Lambda_{\ell}d\Lambda_{r}dl_{\ell}dl^{b}_{\ell r}dl_{r}d\beta_{r}d\zeta_{r} 
\ea
\ee
where $l^b_{\ell,\, r}$ is the length of the effective bond connecting
the last two clusters. 

The Laplace transform of $J$ is considerably simpler to write:
\be
\ba{c}
J{(\Lambda,\Gamma,y)}=\vspace{2mm}\\a_{\Gamma}\l[P{(0,y)}\l(\int\omega{(\beta,\lambda,y)}d\beta\r)^{2}+2\int e^{-2\zeta\lambda}P{(\zeta,y)}d\zeta\omega{(0,\Lambda_{\ell},l_{\ell})}\int\omega{(\beta,\lambda,y)}d\beta\r]=\vspace{2mm}\\
a_{\Gamma}P{(0,y)}\frac{\omega{(\lambda,\Gamma)}^{2}}{(\tau{(y|\Gamma)}+\lambda)}\left(\frac{1}{(\tau{(y|\Gamma)}+\lambda)}+2\frac{1}{(u{(y|\Gamma)}+2\lambda)}\right).  \label{Jeq}
\ea
\ee
\end{widetext}

Eq. (\ref{Jeq}) is one of the main results of this paper; from it we
will  derive the typical and average correlation functions, as well as
information on the distribution of $C_L$. 

%--------------------------------------------------zz x correlations

\section {Exchange Energy and Cross Correlations \label{szsz-sx}}

\subsection{Boundaries and Duality }

The calculation of the $h\sx{0}-h\sx{L}$ correlations in the previous
section  simplified greatly because the operators $\sx{0,\,L}$ considered
were end operators. In this section we calculate expressions for the
end-to-end correlations of the exchange energy density,
$J\sz{}\sz{}$. We consider the special case for which one or both
end-transverse-fields, $h_0$ and $h_L$, are zero, which simplifies the
calculation considerably. We will argue that the universal features
of the correlations will be the same as in the general case with
non-zero end-fields.
                                                                                
The simplifications with vanishing end transverse-fields arise because
this makes the exchange energy be an edge operator in the sense that it is the
first and last energy operator in the Hamiltonian:
\be
\H=-J_{01}\sz{0}\sz{1}-h_1\sx{1}-\ldots-h_{L-1}\sx{L-1}-J_{L-1\,L}\sz{L-1}\sz{L}.
\label{hedgeh}
\ee
And edge bond is the dual of an edge site, and therefore the
edge-to-edge exchange-energy correlations of the ground state of the Hamiltonian in Eq. (\ref{hedgeh})
are duals to the transverse-spin correlations calculated in
Sec. \ref{sx-sx}. This is explained below.

As was explained in the introduction
(\ref{intdual}), the duality transforms a bond to a spin and vice
versa. In the previous section we considered a chain that terminates
with a site that has a finite transverse field on it, $h_0>0$. The
dual of this edge is a chain edge that terminates with a nonzero bond,
$J_{0'1'}=h_0$. The site $1'$ is the dual of the bond $J_{01}$ and
therefore experiences a field $h_{1'}=J_{01}$. 
Since there is no bond $J_{-1~0}$, i.e., $J_{-1~0}=0$, the field on
site $0'$ is zero as well (see Fig. \ref{endfig}).  

In what follows we calculate the end-to-end correlations of the exchange energy
and the cross-correlations between the transverse spin and the
exchange energy. In both cases we will assume that the chain
terminates with the energy operators whose correlations we
calculate (as in Eq. (\ref{hedgeh}) for the exchange-energy correlations). In the case of exchange energy correlations both edges of
the chain we consider will terminate with a vanishing transverse
field, $h_0=h_L=0$. Similarly, when we calculate the edge correlations
between the exchange energy of sites $0$ and $1$,  and the transverse
spin on site $L$, the edge transverse field $h_0$ is set to 0. These
rules allow us to use the dual of the function
$\omega{(\beta,l,\Lambda|\Gamma)}$ which was derived in
Sec. \ref{evsec}. We define the function
$\phi{(\zeta,l,\Lambda|\Gamma)}$ as the dual of
$\omega{(\beta,l,\Lambda|\Gamma)}$. This function will keep track of
the correlations and evolution of the operator $J_{01}\sz{0}\sz{1}$ in the same way
that $\omega{(\beta,l,\Lambda|\Gamma)}$ was used to keep track of the
correlations and evolution of the operator $h_0\sx{0}$.

The calculations carried out in this section assume that one or both
edge transverse fields are zero, but this does not limit the generality of our results for {\it
universal} quantities. We expect that when
the edge transverse-fields are non-zero, the correlations of the last bond in
a chain will only be modified by a non-universal multiplicative factor
from the correlations in the special case with no transverse-field on the end spin.

\subsection{Evolution of Edge Exchange-Energy Operator}

As discussed above, the edge exchange-energy $J_{01}\sz{0}\sz{1}$ is
dual to the and edge transverse field operator
$h_{0}\sx{0}$. Therefore we can obtain the distribution function for
the evolution of the edge exchange energy operator from the results of
Sec. \ref{sx-sx}. 

By making use of the duality (Sec. \ref{intdual}),
we can transform all the results obtained in Sec. $\ref{sx-sx}$ to the
dual chain. As stated above, we define the analog of
$\omega{(\beta,l,\Lambda|\Gamma)}$ to be
$\phi{(\zeta,l,\Lambda|\Gamma)}$:
$\phi{(\zeta,l,\Lambda|\Gamma)}$ keeps track of the bond strength of
the end bond, its length (including the length of the $h=0$ end site),
and the log contribution to the correlation,
$\Lambda$. $\phi{(\zeta,l,\Lambda|\Gamma)}$ is
obtained from the dual of ($\ref{Omega}$):
\be
\ba{c}
\phi{(\zeta,y,\lambda)}=\Phi{(y,\lambda)}\cdot e^{-\zeta\lambda-\zeta u{(y,\Gamma)}}\vspace{2mm}\\
\phi{(y,\lambda)}=e^{(\lambda-\delta)(\Gamma-\Gamma_{I})}\frac{\sinh(\Delta{(y)}\Gamma_{I})}{\sinh(\Delta{(y)}\Gamma)}\frac{2\lambda+\tau{(y,\Gamma_{I})}}{2\lambda+\tau{(y,\Gamma)}}u{(0,\Gamma)}.  \label{phi}
\ea
\ee

\subsection{Exchange Energy Correlations}

The results for the exchange energy correlations are given, by duality,
by Eq. (\ref{Jeq}), with $\delta\rightarrow -\delta$. 
This yields
\be
\ba{c}
J{(\Lambda,\Gamma,y)}=\vspace{2mm}\\
a_{\Gamma}R{(0,y)} \frac{\Phi{(y,\lambda)}^2}{(u{(y|\Gamma)}+\lambda)}\left(\frac{1}{(u{(y|\Gamma)}+\lambda)}+2\frac{1}{(\tau{(y|\Gamma)}+2\lambda)}\right).  \label{Jeq11}
\ea
\ee
Since the results for the exchange-energy correlations are identical
to that of the transverse correlations, we will only analyze the
later.

\subsection{Cross-Correlations --- Last Decimation Step and Final Expression}

In order to obtain the cross correlations we need to combine the
results for the edge transverse-spin flow and exchange-energy
flow. In analogy to Sec.   \ref{sx-sx}, putting together the two flows
happens in the penultinate step of the RG flow. The accumulated
multiplicative factors, along with the couplings of the renormalized
chain just before it is completely decimated, will determine the total
correlations between the transverse-spin and exchange energy.

In contrast to Sec. \ref{sx-sx},  the last needed step of the RG to
obtain the cross correlations involves the transverse spin of one of
the two clusters, and the bond between them:
\be
\ba{l}
C^{x-B}_L=<G|h_0\sx{0}J_{L-1\,L}\sz{L-1}\sz{L}\pgs\vspace{2mm}\\
-<G|h_0\sx{0}\pgs<G|J_{L-1\, L}\sz{L-1}\sz{L}\pgs\vspace{2mm}\\   \label{finalxzz}
\hspace{10mm}=<H|e^{-\Lambda_{\ell}}\sx{\tilde{\ell}}\sz{\tilde{\ell}}\sz{\tilde{r}}e^{-\Lambda_{r}}\pni\vspace{2mm}\\
-<H|e^{-\Lambda_{\ell}}\sx{\tilde{\ell}}\pni<H|\sz{\tilde{\ell}}\sz{\tilde{r}}e^{-\Lambda_{r}}\pni,
\ea
\ee
where $B$ stands for bond. The two possibilities for the last step of the decimation process are the decimation of the bond ($J_{\tilde{\ell}\tilde{r}}$), or of the $\ell$ cluster ($h_{\tilde{\ell}}$). These two processes are dual to each other; hence we only need to consider one of them. Let us consider the site decimation. 

As before, the ground state will be
\be
\pni=|\rightarrow_{\tilde{\ell}}>|\rightarrow_{\tilde{r}}>=\frac{1}{2}(|\uparrow_{\tilde{\ell}}>+|\downarrow_{\tilde{\ell}}>)(|\uparrow_{\tilde{r}}>+|\downarrow_{\tilde{r}}>).
\ee
From the  transformation $S_a=-\frac{J_{\ell r}}{\tilde{h}_\ell}\sz{\tilde{r}}\sy{\tilde{\ell}}$ that induces this decimation, the correlations are found to be:
\be
\ba{c}
C^{x-B}_L e^{\Lambda_{\ell}+\Lambda_{r}}=<H|e^{iS_a}\sx{\tilde{\ell}}e^{-iS_a}e^{iS_a}\sz{\tilde{\ell}}\sz{\tilde{r}}e^{-iS_a}\pni
\vspace{2mm}\\
-<H|e^{iS_a}\sx{\tilde{\ell}}e^{-iS_a}\pni<H|e^{iS_a}\sz{\tilde{\ell}}\sz{\tilde{r}}e^{-iS_a}\pni\vspace{2mm}\\
=<H|(\sx{\tilde{\ell}}-\frac{J_{\ell r}}{\tilde{h}_{\ell}}\sz{\tilde{\ell}}\sz{\tilde{r}})(\sz{\tilde{\ell}}\sz{\tilde{r}}+\frac{J_{\ell r}}{\tilde{h}_{\ell}}\sx{\tilde{\ell}})\pni\vspace{2mm}\\
-<H|(\sx{\tilde{\ell}}-\frac{J_{\ell r}}{\tilde{h}_{\ell}}\sz{\tilde{\ell}}\sz{\tilde{r}})\pni<H|(\sz{\tilde{\ell}}\sz{\tilde{r}}+\frac{J_{\ell r}}{\tilde{h}_{\ell}}\sx{\tilde{\ell}})\pni>\vspace{2mm}\\
=-\frac{J_{\tilde{\ell}\tilde{r}}}{\tilde{h}_{\ell}}=-e^{-\zeta_{\tilde{\ell}\tilde{r}}}.
\ea  \label{negzzx}
\ee
Thus the cluster decimation process yields
\be
\Lambda=\Lambda_{\ell}+\Lambda_{r}+\zeta_{\tilde{\ell}\tilde{r}}.
\ee
By duality, the bond decimation process yields
\be
\Lambda=\Lambda_{\ell}+\Lambda_{r}+\beta_{\tilde{\ell}}.
\ee

Following the reasoning that led to Eqs. ($\ref{Jps},~\ref{Jeq}$) we get:
\be
J{(\lambda,\Gamma|y)}=
a_{\Gamma}\omega{(\lambda,\Gamma)}\phi{(\lambda,\Gamma)}\left(\frac{1}{\tau{(y|\Gamma)}+\lambda}+\frac{1}{u{(y|\Gamma)}+\lambda}\right).  \label{Jeqxzz}
\ee
This is the second main result in this paper, and it is analogous to
Eq. (\ref{Jeq}). Here we must bear in mind that the correlations
obtained here are negative (see Eq. \ref{negzzx}). This is to be expected, since the two operators,
$\sx{}$ and $\sz{}\sz{}$, try to impose competing orders; One tends to disorder the system and the other to order it.

%---------------------------------------------results

\section{Results \label{results}}

The above results (\ref{Jeq},~\ref{Jeqxzz}) in principal allow the
calculation of the distribution function for the log-correlations,
$\Lambda=-\log(C_L)$, of long finite chains.  In the following sections
we calculate the average correlations, $\overline{C_L}$, and the
distribution, $f(\Lambda|L)$, for all $\delta$.

\subsection{Average $h\sx{} - h\sx{}$ and $J\sz{}\sz{}-J\sz{}\sz{}$ Correlations}

\paragraph{Derivation of the Average.} 

In this section we derive the {\it average} x-x correlations. The
BB correlations of the exchange energy are obtained from the x-x
correlations upon the transformation $\delta\rightarrow -\delta$. 

In order to obtain $\overline{C^{xx}_{L}}$, we begin with Eq. (\ref{Jeq}) in the following form:
\begin{widetext}
\be
\ba{c}
J{(\lambda,\Gamma|y)}=\frac{\tau{(0,\Gamma)}}{u{(0,\Gamma)}}\frac{\sinh^{2}(\Delta\Gamma_{I})}{\sinh(\Delta\Gamma)}\Delta\frac{(2\lambda-\delta+\Delta \coth(\Delta\Gamma_{I}))^{2}}{(2\delta-\lambda)^{2}}e^{-(2\lambda+3\delta)\Gamma+2(\lambda+\delta)\Gamma_{I}}\vspace{2mm}\\
\left[ \frac{1}{[(\delta+\lambda)\sinh(\Delta\Gamma)+\Delta\cosh(\Delta\Gamma)]^{2}}
+\frac{2(2\delta-\lambda)\sinh(\Delta\Gamma)}{[(2\lambda-\delta)\sinh(\Delta\Gamma)+\Delta\cosh(\Delta\Gamma)]^{3}}
-\frac{1}{[(2\lambda+\delta)\sinh(\Delta \Gamma)+\Delta\cosh(\Delta\Gamma)]^{2}} \right].
\ea
\label{Jeq1}
\ee
First, we perform an inverse Laplace transform in $y$ to recover the length dependence:
\be
\ba{c}
J{(\lambda,\Gamma|L)}=\sum\limits_{n=1}^{\infty}(-1)^n\frac{\tau{(0,\Gamma)}}{u{(0,\Gamma)}}\frac{((2\lambda-\delta)\Gamma_I+1)^2}{(2\delta-\lambda)^2}
e^{-(2\lambda+3\delta)\Gamma+2(\lambda+\delta)\Gamma_I}e^{ -\left(\delta^2+\left(\frac{n\pi}{\Gamma}\right)^2\right)L}\vspace{2mm}\\
\left(\frac{4L(n\pi)^4}{\Gamma^6(\lambda+\delta)\left(1+\frac{\Gamma^3(\lambda+\delta)}{2(n\pi)^2L}\right)}e^{\frac{2L(n\pi)^2}{(\lambda+\delta)\Gamma^3}}
+\frac{8L^2(n\pi)^6(2\delta-\lambda)}{\Gamma^9(2\lambda-\delta)^3} e^{\frac{2L(n\pi)^2}{(2\lambda-\delta)\Gamma^3}}
-\frac{4L(n\pi)^4}{\Gamma^6(2\lambda-\delta)\left(1+\frac{\Gamma^3(2\lambda-\delta)}{2(n\pi)^2L}\right)}e^{\frac{2L(n\pi)^2}{(2\lambda- \delta)\Gamma^3}}\right).
\ea
\ee
\end{widetext}
This  is obtained by approximating the roots of 
\be
a\sinh(\Delta\Gamma)+\Delta\cosh(\Delta\Gamma) \label{den}
\ee
with $a\approx 1$, by
$y_{n}=-\delta^{2}-\left(\frac{n\pi}{\Gamma}\right)^2\left(1-\frac{2}{\Gamma a}\right)$
and expanding (\ref{den}) around these roots:
\be
a\sinh(\Delta\Gamma)+\Delta\cosh(\Delta\Gamma)\approx\sum\limits_{n=1}^{\infty}(-1)^{n+1}i(y-y_{n})\frac{a\Gamma^{2}}{2n\pi}.
\label{yroots}
\ee
The roots in Eq. (\ref{yroots}) are given as an expansion in powers of
$1/\Gamma$. Since we are interested in $\Gamma\gg 1$ we are content
with only the first two terms; in fact, as can be seen by the following
Eq. (\ref{sp}), only the first nonvanishing power of $1/\Gamma$
contributes to the average correlations. In addition,
Eq. (\ref{yroots}) is only valid for $y_n\ll 1$, i.e., for
$n\pi<\Gamma$. But since we are
interested in the large length behavior of the correlations, we can
restrict our calculation to small values of $n$, as they give the
slowest decaying term in the correlations.

The desired result is obtained by performing the $\Gamma$
integral. This integral is dominated by the large exponent in
\be
e^{-\left(\frac{n\pi}{\Gamma}\right)^{2}L-(2\lambda+\delta)\Gamma\,(+O{(\log(\Gamma))}+O{(L\Gamma^{-3})})}=e^{g{(\Gamma)}}
\label{sp}
\ee
which has a saddle point at $\Gamma_{S}=\left(\frac{2(n\pi)^{2}L}{2\lambda+\delta}\right)^{1/3}$. The exponential dependence then becomes
\be
e^{g{(\Gamma)}}\approx e^{g{(\Gamma_{S})}-\frac{1}{2}3\left(\frac{(2\lambda+\delta)^{4}}{2(n\pi)^{2}L}\right)^{1/3}(\Gamma-\Gamma_S)^{2}} 
\ee
and the saddle point integration yields
\begin{widetext}
\be
\ba{c}
J{(\lambda,L)}\approx\sum\limits_{n=1}^{\infty}(-1)^{n+1}2^{2/3}\sqrt{\frac{\pi}{3}}(n\pi)^{1/3}\frac{1}{L^{5/6}}e^{-\delta^{2}L-3L^{1/3}(n\pi)^{2/3}(\lambda+\delta/2)^{2/3}}\vspace{2mm}\\
\times e^{2\Gamma_{I}(\lambda+\delta)}\left(\frac{2\lambda\Gamma_{I}+1}{\lambda-2\delta}\right)^{2}\frac{1}{(2\lambda+\delta)^{2/3}}\vspace{2mm}\\
\times \left(e^{2-\frac{\delta}{\lambda+\delta}}\frac{(2\lambda+\delta)^3}{(\lambda+\delta)(3\lambda+2\delta)}+e^{1-\frac{2\delta}{2\lambda-\delta}}\frac{(2\delta-\lambda)(2\lambda+\delta)^2}{(2\lambda-\delta)^2}-e^{2-\frac{\delta}{\lambda+\delta}}\frac{(2\lambda+\delta)^3}{(4\lambda)(2\lambda-\delta)}\right).  \label{xxfres}
\ea
\ee
\end{widetext}
This result is valid for the critical regime ($\frac{1}{\delta^2}\gg L$) when $\frac{1}{L^{1/3}}\ll\lambda\ll L$, and away from the critical regime ($\frac{1}{\delta^2}\ll L$) away from $\lambda=0$. To get the equivalent expression for $\lambda\Rightarrow 0$, we  need to be more careful with the third term of equation (\ref{xxfres}) and get the next order corrections.
Note that there is no singularity in this expression at
$\lambda=2\delta$; this will have implications for the off-critical large $L$ behavior.

\paragraph{Result.}
To get the final result for $\overline{C^{xx}_{L}}$ all that remains is to set $\lambda\Rightarrow 1$. Neglecting terms suppressed by factors of $\delta\ll 1$ we obtain
\be
\overline{C^{xx}_{L}}\approx A^{xx}_{0}\frac{1}{L^{5/6}}e^{-\delta^{2}L-3L^{1/3}(\pi)^{2/3}(1+\delta/2)^{2/3}},
\label{favexx}
\ee
Also, for the exchange-energy correlations we obtain
\be
\overline{C^{BB}_L}\approx A^{zz}_{0}\frac{1}{L^{5/6}}e^{-\delta^{2}L-3L^{1/3}(\pi)^{2/3}(1-\delta/2)^{2/3}},
\label{favezz}
\ee
where $A^{xx}_{0}$ and $A^{zz}_{0}$ are non-universal coefficients.

When the chain is not critical, we notice that the exponential decay is controlled by the same correlation length, $\xi=\frac{1}{\delta^{2}}$, as  the order parameter correlation function. 

%---------------------------------------------------------------------------x-B derivation+results
\subsection {Average Cross Correlations}

\paragraph{Derivation of the Average.}

In complete analogy with the derivation of the previous section, we proceed from Eq. (\ref{Jeqxzz}) in the following form:
\begin{widetext}
\be
\ba{c}
J{(\lambda,\Gamma|y)}\approx\frac{((2\lambda\Gamma_{I}+1)^2-(\delta\Gamma_{I})^2)}{\lambda}e^{-2\lambda(\Gamma-\Gamma_{I})}\frac{\Delta^2}{\sinh(\Delta\Gamma)}\vspace{2mm}\\
\times \left(\frac{1}{(\lambda+2\delta)}\frac{1}{((\lambda-\delta)\sinh(\Delta\Gamma)+\Delta\cosh(\Delta\Gamma))}+\frac{1}{(\lambda-2\delta)}\frac{1}{((\lambda+\delta)\sinh(\Delta\Gamma)+\Delta\cosh(\Delta\Gamma))}\right.\vspace{2mm}\\
\left.+\frac{(\lambda-\delta)}{\delta(2\delta-\lambda)}\frac{1}{((2\lambda-\delta)\sinh(\Delta\Gamma)+\Delta\cosh(\Delta\Gamma))}+\frac{(\lambda+\delta)}{\delta(2\delta+\lambda)}\frac{1}{((2\lambda+\delta)\sinh(\Delta\Gamma)+\Delta\cosh(\Delta\Gamma))}\right).
\ea
\ee
Performing an inverse Laplace transform in y and then performing a saddle point integration in $\Gamma$, we find:
\be
\ba{c}
J{(\lambda|L)}\approx \sum\limits_{n=1}^{\infty} e^{2\lambda\Gamma_I}\frac{((2\lambda\Gamma_{I}+1)^2-(\delta\Gamma_{I})^2)}{\lambda^{2/3}}2\sqrt{\frac{\pi}{3}}(n\pi)^{1/3}\frac{1}{L^{5/6}}e^{-\delta^{2}L-3L^{1/3}(n\pi)^{2/3}\lambda^{2/3}}\vspace{2mm}\\
\left(\frac{e^{2+\frac{2\delta}{\lambda-\delta}}}{\lambda-2\delta}+\frac{e^{2-\frac{2\delta}{\lambda+\delta}}}{\lambda+2\delta}-\frac{(\lambda-\delta)e^{1+\frac{\delta}{2\lambda-\delta}}}{\delta(\lambda-2\delta)}+\frac{(\lambda+\delta)e^{1-\frac{\delta}{2\lambda+\delta}}}{\delta(\lambda+2\delta)}\right).
\ea
\ee
This result, as well as Eq. (\ref{xxfres}), is valid for $\frac{1}{L^{1/3}}\ll\lambda\ll L$ for all small or zero $\delta$. 
\end{widetext}

\paragraph{Result.}
The average x-B correlation is  obtained from the above by setting $\lambda=1$:
\be
\overline{C^{x-B}_{L}}\approx -A^{x-B}_{0}\frac{1}{L^{5/6}}e^{-\delta^{2}L-3L^{1/3}(\pi)^{2/3}},
\ee
which is almost the same as Eq. (\ref{favexx}) but the $\delta$
dependence of the above result is strictly symmetric with respect to $\delta$, 
as expected for an object that is self-dual.

%-------------------------------------------------typical correlation

\subsection {Typical Correlations}

One of the striking features of random quantum systems is that
typical correlations are usually very different from average
correlations. Average correlation functions can be, as here,
dominated by samples (or spatial regions)  with anomalously strong
correlations. The typical correlations are much smaller and decay
faster with distance. Indeed, for the random Ising chain, the typical
correlations hold for {\it almost all long-but-finite samples}. The
average end-to-end correlations are dominated by extremely rare
samples.

In the off-critical regime the typical correlations are well characterized by the average log-correlations
\begin{center}
$C_{typical}=e^{-\overline{\log(C)}}=e^{-\overline{\Lambda}}$.
\end{center}
However, at the critical point $ -\overline{\log(C)}=
-\overline{\Lambda}$ is of the same {\it order} as the logarithm of
the typical correlations,  but the typical correlations will also have
a very wide spread. More precisely, there is a random
proportionality Constant relating the log of the correlations to its
average; this constant is random and
widely varying. 
We will first investigate the typical correlations at the critical point, and then consider the off critical regime.

%-------------------------------------------------------typical correlations at the critical point

\paragraph{Typical Correlations at the critical point.}

The average log-correlations $\overline{\Lambda}$ are easily found at
the critical point. Going back to Eq. (\ref{Jeq}) and setting
$\delta=0$, we see that the Laplace transform in $L$ of the x-x
log-correlations is given by
\begin{widetext}
\be
\ba{c}
\overline{\Lambda}^{xx}(y)=-\left.\int\limits_{\Gamma_I}^{\infty}\der{J{(\lambda,\Gamma,y)}}{\lambda}\right|_{\lambda\Rightarrow 0}d\Gamma\vspace{2mm}\\
=\int\limits_{\Gamma_I}^{\infty}\frac{2\tanh(\Gamma\sqrt{y})(3(\Gamma-3\Gamma_I)\sqrt{y}+10\tanh(\Gamma\sqrt{y}))}{y\cosh(\Gamma\sqrt{y})^3}d\Gamma\vspace{2mm}\\ \label {derr}
=\frac{2}{y^{3/2}}\left(3\int\limits_{\sqrt{y}\Gamma_I}^{\infty}\frac{\sinh(x)}{\cosh^4(x)}xdx+10\int\limits_{\sqrt{y}\Gamma_I}^{\infty}\frac{\sinh^2(x)}{\cosh^5(x)}dx-9\sqrt{y}\Gamma_I\int\limits_{\sqrt{y}\Gamma_I}^{\infty}\frac{\sinh(x)}{\cosh^4(x)}dx\right)\vspace{2mm}\\
\approx\frac{7\pi}{4y^{3/2}}+O\left(\frac{1}{y}\right).
\ea
\ee
\end{widetext}
By performing the inverse Laplace transform we get
\be
\ba{c}
\overline{\Lambda}_{L}^{xx}\approx\frac{7\sqrt{\pi}}{4}\sqrt{L}+O(1)\vspace{2mm}\\ 
\label{avelxx}
\approx 3.1 \sqrt{L}.
\ea
\ee

The result in Eq. (\ref{avelxx}) should be compared with the critical behavior of the average correlation:
\be
\log\left(\overline{e^{-\Lambda}}\right)\approx -3\pi^{2/3}L^{1/3}.
\ee
The typical correlations decay as $\sim e^{-k\sqrt{L}}$ with $k$ of order unity but random with a computable universal distribution. But the average correlation function is $\sim e^{-c'L^{1/3}}$, which decays much more slowly. As claimed above, this means that realizations of the quenched randomness that have an exponentially low probability dominate the average. 
 
By differentiating (\ref{derr}) once more with respect to $\lambda$, we get $\overline{\l(\Lambda_{(y)}^{xx}\r)^2}$. Using that we get for the standard deviation of $\Lambda^{xx}$:
\be
\sqrt{\overline{\l(\Lambda^{xx}\r)^2-\overline{\Lambda^{xx}}^2}}\approx 5.6\sqrt{L}.
\ee
The distribution of $\Lambda/\sqrt{L}$ is thus indeed non-trivial for long critical chains.

For the cross correlation function, $J\sz{}\sz{} - h\sx{}$, the result we get for the average  log-correlation is (by a similar calculation)
\be
\begin{array}{c}
\overline{\Lambda^{x-B}_L}\approx\frac{16}{3\sqrt{\pi}}\sqrt{L}+O(1)\vspace{2mm}\\
\approx 3.0\sqrt{L},\vspace{2mm}\\  
\label{avelzzx}
\sqrt{\overline{\l(\Lambda^{x-B}\r)^2-\overline{\Lambda^{x-B}}^2}}\approx 5.4\sqrt{L}.
\end{array}
\ee
Note the similar --- but not identical --- behavior of the two results (\ref{avelzzx}) and (\ref{avelxx}). 

\paragraph{Off Critical x-x Correlations.  \label{octypical}}

To investigate the x-x energy correlations in the off-critical regime, we pursue a different course of action. Instead of setting $\lambda$ to 1, we  invert the Laplace transform  with respect to $\lambda$ in expression (\ref{Jeq}) and obtain $J{(\Lambda,\Gamma|y)}$ in terms of $\Lambda$.

Equation (\ref{Jeq}) can be written in the following form:
\begin{widetext}
\be
\ba{c}
J{(\lambda,\Gamma|y)}=\frac{\tau{(0,\Gamma)}}{u{(0,\Gamma)}}\frac{\sinh^{2}(\Delta\Gamma_{I})}{\sinh(\Delta\Gamma)^3}\
\Delta(2\lambda-\delta+\Delta \coth(\Delta\Gamma_{I}))^{2}e^{-2\lambda(\Gamma-\Gamma_I)} e^{+2\delta\Gamma_{I}-3\delta \Gamma}\vspace{2mm}\\
\cdot \left[ \frac{1}{(3\delta+\Delta\coth(\Delta\Gamma))^{2}(\delta+\lambda+\Delta\coth(\Delta\Gamma))^2}\right.
+\frac{2}{(3\delta+\Delta\coth(\Delta\Gamma))^{3}(\delta+\lambda+\Delta\coth(\Delta\Gamma))}\vspace{2mm}\\
+\frac{4}{(3\delta+\Delta\coth(\Delta\Gamma))(2\lambda-\delta+\Delta\coth(\Delta\Gamma))^3}
\left.-\frac{4}{(3\delta+\Delta\coth(\Delta\Gamma))^3(2\lambda-\delta+\Delta\coth(\Delta\Gamma))} \right].
\ea
\ee
This form of $J{(\lambda,\Gamma|y)}$ lends itself to inverting the Laplace transform and recovering the $\Lambda$ dependence. This gives (neglecting  $\Gamma_I$, as before)
\be
\ba{c}
J{(\Lambda,\Gamma|y)}\approx\frac{\tau{(0,\Gamma)}}{u{(0,\Gamma)}}\frac{1}{\sinh(\Delta\Gamma)^3}
\Delta^3 e^{+2\delta\Gamma_{I}-3\delta \Gamma}
\Theta{(\Lambda-2(\Gamma-\Gamma_I))}\vspace{2mm}\\
\left[\left( \frac{(\Lambda-2(\Gamma-\Gamma_I))}{(3\delta+\Delta\coth(\Delta\Gamma))^{2}}
+\frac{2}{(3\delta+\Delta\coth(\Delta\Gamma))^{3}}\right)e^{-(\Lambda-2(\Gamma-\Gamma_I))(\delta+\Delta \coth(\Delta\Gamma))}\right.\vspace{2mm}\\
+\left(\frac{2(\Lambda-2(\Gamma-\Gamma_I))^2}{(3\delta+\Delta\coth(\Delta\Gamma))}-\frac{4}{(3\delta+\Delta\coth(\Delta\Gamma))^3}\right)\left.e^{-\frac{1}{2}(\Lambda-2(\Gamma-\Gamma_I))(-\delta+\Delta \coth(\Delta\Gamma))} \right], \label{lambdainv1}
\ea
\ee
with $\Theta$ the Heaviside step function. Off critical, for long enough chains, specifically with, $L\gg\xi\approx\frac{1}{\delta^2}$ and the concomitant log-energy scale $\Gamma\delta\gg 1$, one can expand 
\be
\Delta\approx\delta+\frac{y}{2\delta}
\ee and hence obtain
\be
\ba{c}
J{(\Lambda,\Gamma|y)}\approx|\delta|^3e^{+2\delta\Gamma_{I}-(\delta+3|\delta|) \Gamma}
\Theta_{(\Lambda-2\Gamma)}\vspace{2mm}\\
\left[\left( \frac{(\Lambda-2(\Gamma-\Gamma_I))}{(4\delta)^{2}}
+\frac{2}{(4\delta)^{3}}\right)e^{-(\Lambda-2(\Gamma-\Gamma_I))(\delta+|\delta|)}e^{-\frac{y}{2|\delta|}(\Lambda-2(\Gamma-\Gamma_I))}\right.\vspace{2mm}\\
+\left(\frac{2(\Lambda-2(\Gamma-\Gamma_I))^2}{(4\delta)}-\frac{4}{(4\delta)^3}\right)\left.e^{-\frac{1}{2}(\Lambda-2(\Gamma-\Gamma_I))(-\delta+|\delta|)}e^{-\frac{y}{4|\delta|}(\Lambda-2(\Gamma-\Gamma_I))} \right].  \label{lambdainv}
\ea
\ee
From the simple form of the $y$ dependence in this limit, one can
invert the Laplace transform  by inspection to obtain the $L$ dependence. 

In the paramagnetic phase ($\delta>0$), $J(\Lambda,\Gamma|L)$ is
sharply peaked for long chains at $2(\Gamma-\Gamma_I)+4\delta L$. In
the ferromagnetic phase ($\delta<0$), it is instead sharply peaked at
at $2(\Gamma-\Gamma_I)+2|\delta| L$. Integrating over $\Gamma$ gives
exponential decay in $L$. Thus the distributions of the end-to-end transverse field log-correlations  of long off-critical samples will have the form :
\be
f^{xx}{(\Lambda,L)}=\int\limits_{\Gamma_I}^{\infty}J{(\Lambda,\Gamma|L)}d\Gamma\sim\left\{
\ba{lc}
e^{-2\delta(\Lambda-4\delta L)}\Theta{(\Lambda-4\delta L)} &	\delta>0\vspace{2mm}\\
e^{-|\delta|(\Lambda-2|\delta|L)}\Theta{(\Lambda-2|\delta| L)} 	&	\delta<0
\ea
\right. .
\ee
[Recall that $\eta$ is the Heaviside step function.]
This behavior, with exponential decay of almost all samples with a characteristic length that is much shorter than the correlation length is similar to that of the order parameter correlations in the paramagnetic phase as discussed in Sec. \ref{spinintro} of the Introduction.
\end{widetext}

\paragraph{Off Critical x-B Correlations.} The same analysis can be applied to the x-B correlation function. The mathematical expressions are simpler, but the result is more interesting. Since this correlation function is symmetric with respect to $\delta$, we can choose to carry out the analysis in the paramagnetic phase, $\delta>0$. Using the same simplifying limit as before ($\Gamma_I\Rightarrow 0,\,\Gamma\delta\gg 1,\,y\ll\delta^2$), and keeping only the dominating terms in the disordered phase, Eq. (\ref{Jeqxzz}) becomes
\be
J{(\lambda,\Gamma|y)}\approx e^{-2\lambda(\Gamma-\Gamma_I)-2\delta\Gamma}\delta^2
\left(\frac{1/y}{\lambda-\frac{y}{4\delta}}-\frac{1/y}{\lambda-\frac{y}{2\delta}}\right).
\ee
The inverse Laplace transform in $\lambda$ and $y$ of this leads to
\be
\ba{c}
J{(\lambda,\Gamma|y)}\approx
e^{-2\delta\Gamma}\delta^2\vspace{2mm}\\
\l(\Theta{\l[\Lambda-2(\Gamma-\Gamma_I)-2\delta L\r]}-\Theta{\l[\Lambda-2(\Gamma-\Gamma_I)-4\delta L\r]}\r).
\ea
\ee
By integrating over $\Gamma$ we get for the distribution of the
end-to-end log cross-correlations,  $\Lambda=\Lambda^{x-B}_L$, of long off-critical chains,
\begin{widetext}
\be
f^{x-B}_{(\Lambda,L)}=\int\limits_{\Gamma_I}^{\infty}J{(\Lambda,\Gamma|L)}d\Gamma\sim\left\{
\ba{lc}
0	&	\Lambda<2|\delta| L\vspace{2mm}\\
1-e^{-\delta(\Lambda-2|\delta| L)} 	& 	2|\delta| L<\Lambda<4|\delta| L \vspace{2mm}\\
(1-e^{-2\delta^2 L})e^{-|\delta|(\Lambda-4|\delta| L)}	& \Lambda>4|\delta| L
\ea
\right. .
\ee
\end{widetext}
Most long samples will have $2\delta L<\Lambda<4\delta L$; remembering that this expression is valid only for $L\delta^2\gg1$, we see that the distribution will be roughly constant in this range. For  larger $\Lambda$, the distribution decays exponentially. 

In the ferromagnetic phase, the same result for the cross correlations will obtain with $\delta$ replaced by $|\delta|$. The behavior in this phase contrasts with that of  the x-x correlations whose distribution of $\Lambda$ is peaked near $2|\delta| L$ and thus are typically stronger than the cross correlations.

%AAAAAAAAAAAAAAAAAAAAAAAAAAAAAAAAAAAAAAAAAAAAAAAAAAAAAAAAAAAAAAAAAAAAAAAAAAAAAAAAAAAAAAAAAaa

\subsection{Energy - Energy Correlations and the Energy Gap \label{rell}}

Looking  at the earlier results for the average of the energy gap,
$\Delta E$, of finite chains, \cite{DSF98} we observe a strong
resemblance to the results obtained here for the average E - E
correlations. In particular Eq. (60) in Ref. [\onlinecite{DSF98}], 
\be
\overline{\Delta E}\sim L^{1/6}e^{-\frac{3}{2}\left(\frac{\pi^2L}{2}\right)^{1/3}},
\ee
gives the average gap at the critical point $\delta=0$. We see that
\be
2\log\overline{\Delta E}\approx  \log \overline{C^{EE}_L}
\ee
Some relation between the gap and the energy correlations is to be
expected, but the behavior of the two quantities is surprising in the
degree of the similarity. We will see that this relationship between
the gap and the energy correlations arises in the structure
of the RG flow.

In Eq. (\ref{lambdainv}), the Heaviside function
$\Theta{(\Lambda-2(\Gamma-\Gamma_I))}$ implies that
$-\log(C^{\Gamma}_{(L)})>2(\Gamma-\Gamma_I)$. Since the appropriate
$\Gamma$ at which the decimation establishing this correlation occurs
is $\Gamma = -\log \Delta E$, this implies that 
 \be
 C^{xx}_L<\left(\frac{\Delta E}{\Omega_I}\right)^2 \ .
\ee
To understand this inequality,  consider the correlation coefficient
$e^{-\Lambda}$ associated with an end site, and compare this to the
transverse field $\tilde{h}$ on the end spin cluster. The strongest
correlations will occur if the chain undergoes a series of {\it bond}
decimations. Looking at Fig. (\ref{endJfig}) one can see that in this
case the  evolution of $e^{-\Lambda}$ and $\tilde{h}$ are exactly the
same; in each end-bond decimation they acquire a factor of the
$\frac{h_0}{J_{01}}$ at that scale. The strongest correlations
dominate the average correlations. Therefore chains in which the
energy-correlations and the gap are strongly correlated also dominate
the average correlations.

\section{Conclusions \label{conclusions}}

In this paper we have investigated the various contributions to the
end-to-end  energy correlations of random quantum Ising chains in  the
universal regime of long chains near the quantum phase transition.  In
principal, the main result obtained here, the Laplace transform  of
the distribution of the logarithm of the correlation functions
(\ref{Jeq}, \ref{Jeqxzz}), can be used to obtain  the complete
distributions in the scaling limit. We have explicitly computed
the average and typical  correlations, as well as some other
aspects of the distributions, in various limits.  The average
correlations are dominated by exponentially (in the chain length) rare
samples. Nevertheless, they still decay as
$e^{-CL^{1/3}}$ at the critical point. This is in contrast to the power law
decay of the average order parameter correlations.  
The various components of the energy correlations are qualitatively
similar, although their distributions differ. The cross correlations
between the  ordering operator, $J\sigma^z\sigma^z$, and disordering
transverse-field operator, $h\sigma^x$, are negative because of their competing effects;
they are also self-dual. 

The average correlations in the off-critical regime decay with the
same correlation length: $\xi\sim\frac{1}{\delta^2}$, as the order
parameter correlations. The typical correlations, however, decay much
faster, and have the same correlation-length exponent, $\tilde{\nu}=1$, as the
pure system. In Sec. (\ref{octypical}) it was shown that  the distribution
of $\sx{0}\sx{L}$ is strongly peaked near $\exp(-4\delta L)$ for
$\delta>0$ and near $\exp(-2|\delta|L)$ for $\delta<0$, indicating a
surprising asymmetry  between the two phases.  This asymmetry is a
result of the difference between the ordering and disordering
components of the energy density and their behavior in the
corresponding phases. At the critical point, the typical correlations
decay as $\sim e^{-K L^{1/2}}$ with $K$ a random variable; this is
similar to the typical order parameter correlations.

The behavior of the end-to-end energy correlations turn out to be
similar to that of  the energy gap. We explain this in terms of  the
rare realizations of the quenched randomness that dominate the
averages: these are such that the gap and the correlation function
involve essentially the same product of ratios of $J$'s to $h$'s.

Unlike previous RG calculations of properties of the random
transverse-field Ising model, the energy correlations required the
development of a formalism that goes beyond second order perturbation
theory. Performing the RG transformation by unitary transformations
proved to be a useful tool that  allows one to follow readily the
evolution of effective operators. 
Here we have focussed on end-to-end correlations because these are far
simpler to handle analytically: correlations in the bulk of the chain
involve effective operators on both sides of the objects of interest
and are much harder to deal with. Nevertheless, they could be computed
by keeping track of the needed distributions numerically.

The unitary transformation RG method can also be applied to the
calculation of correlations in imaginary time. In this case, primarily
average quantities have been calculated thus far (for instance, see Refs. 
[\onlinecite{kedar,mdh,igloi}]) but progress on distributions should be possible
utilizing the procedure described here (Sec. \ref{formalism}). 
The present method may also be applied to other models both in one
dimension, and, numerically, in higher dimensions.

%In 2D the prospects of analytically deriving distributions of
%correlation functions remain grim. However, in some cases the use of
%the unitary transformations flow may make a 2D or higher dimensional
%problem amenable to 1D methods, by essentially reducing its
%dimensionality. A simple example of that is a 2D nearest neighbor
%random hopping model with:
%\be
%\H_{random-hopping}=\sum\limits_{<ij>}t_{ij}(\hat{c}^{\dagger}_i \hat{c}_j+\hat{c}^{\dagger}_j \hat{c}_i)
%\ee
%If we proceed by picking out a strong bond, e.g. $t_{01}$ we can use
%$S=\frac{t_{12}}{t_01}\left(\hat{c}^{\dagger}_0
%\hat{c}_1-\hat{c}^{\dagger}_0 \hat{c}_2\right)$ to eliminate the
%$t_{12}$. By applying this procedure to a path of strong bonds, we
%are able to disconnect this path from the rest of the chain even
%though the wave function on this path is still not determined. After
%the path is disconnected, we can proceed to apply 1D RG to it. A path
%connecting two operators, may lead to a distribution function of
%their correlation function.
%<<< I COULD NOT REALLY UNDERSTAND THE ABOVE PARAGRAPH (CUT BY  \%) AND I AM NOT SURE IT IS CORRECT >>>

\acknowledgments{This work has been supported in part by the National
  Science Foundation via DMR-9976621, DMR-9809363, and the MRSEC at
  Harvard.  One of us (GR) was supported in part by a Harvard Merit
  Fellowship.} 

%--------------------------------------------------------------appendix

\appendix 
\section{Effects of End Operators \label{appA}}

When calculating the energy correlation functions we kept terms that
were third order or higher in the perturbation expansion. A conccern
to the validity of our treatment is that terms that were produced as
fourth or fifth order terms at an early stage of the RG flow,
$\Gamma_1$, become more relevant than terms that we kept that are
produced at a later stage of the RG flow, $\Gamma_2>\Gamma_1$.  
In this Appendix we verify that end operator effects that we
excluded in the text can not give rise to leading order contributions to
the various computed energy correlations.

In addition, The flow of the operator $\sx{0}$ when the end spin is decimated, will include a term $\sz{0}\sz{1}$. 
Naively, as this is an order --- rather than a disorder --- energy
operator, it may make the disordered and ordered phases look
indistinguishable as far as the transverse spin correlations are
concerned. We will show here that this is {\it not} the case, and
although these extraneous operators do appear, their contribution is
at best subdominant.

\subsection{End Site Decimation}

In order to prove the above claims, we need to investigate the additional operators that arise when decimating ends. 
Revisiting the process of decimating an end-site, we  consider the
flow of the operator $h_0\sx{0}$. The following table describes the
series of transformation, and the effective operators that contribute
to $h_0\sx{0}$:
\begin{widetext}
\be
\ba{ccc}
& \underline{S}  &  \underline{h_0\sx{0}}\vspace{2mm}\\
S_a:\,\,& -\frac{J_{01}}{2h_0}\sy{0}\sz{1} 			&	-J_{01}\sz{0}\sz{1} \vspace{2mm}\\
S_b:\,\,&\frac{J_{01}h_1}{2h^2_0}\sz{0}\sy{1} 			&	-\frac{J_{01}h_1}{h_0}\sy{0}\sy{1}+\frac{J_{01}^2h_1}{h_0^2}\sx{1} \vspace{2mm}\\
S_c:\,\,&-\frac{J_{01}J_{12}h_1}{2h^3_0}\sy{0}\sx{1}\sz{2} 	&	-\frac{J_{01}J_{12}h_1}{h^2_0}\sz{0}\sx{1}\sz{2} \vspace{2mm}\\
S_d:\,\,&-\frac{J_{01}J_{12}h_1h_2}{2h^4_0}\sz{0}\sx{1}\sy{2} \ \ \ \ 	&	\frac{J_{01}^2J_{12}h^2_1}{h^4_0}\sz{1}\sz{2}+\frac{J_{01}^2}{h_0^4}J_{12}h_1h_2\sy{1}\sy{2}. \label{sxflow}
\ea 
\ee
\end{widetext}
The fourth line and the second line contain the terms that are most
likely to give a leading contribution to the correlations; these
operators have a non-vanishing expectation value in the ground state
of the decimated part. Keeping them all we have
\be
h_0\sx{0}\Rightarrow \frac{J_{01}^2h_1}{h_0^2}\sx{1}+\frac{J_{01}^2}{h_0^4}J_{12}h^2_1\sz{1}\sz{2}+\frac{J_{01}^2}{h_0^4}J_{12}h_1h_2\sy{1}\sy{2}.
\label{app2}
\ee

Another process that we need to consider is a bond decimation  close
to the end. If we decimate the first bond, $-J_{01}\sz{0}\sz{1}$, then
$h_0\sx{0}$ becomes
\be
\ba{c}
h_0\sx{0}\Rightarrow -\frac{h_{0}h_1}{J_{01}}\sy{0}\sy{1}+\frac{h_0h_1}{J^3_{01}}J_{12}h_2\sy{0}\sx{1}\sy{2}=\vspace{2mm}\\
h_{(01)}\sx{(01)}+\frac{1}{\Omega^2}h_{(01)}h_{2}J_{(01)\,2}\sy{(01)}\sy{2},
\ea
\ee
where the parentheses signify effective spin clusters.
Going further away from the end, a decimation of the second site in the chain, $-h_1\sx{1}$, will give
\be
\ba{c}
h_0\sx{0}\Rightarrow h_0\sx{0}+\frac{1}{h_1^2}\frac{J_{01}J_{12}}{h_1}h_0h_2\sy{0}\sx{1}\sy{2}=\vspace{2mm}\\
h_0\sx{0}+\frac{1}{h_1^2}J_{02}h_0h_2\sy{0}\sy{2},
\ea
\ee
with site 1 eliminated. Note that site 0 is still the first site but site 2 is now the next.

From the above processes a pattern emerges. The fifth-order
perturbation calculation above produces two dangerous operators:
\be
\sz{0}\sz{1},\,\sy{0}\sy{1}.
\label{dop}
\ee
We need to verify that these operators do not produce leading-order
correlations. The first step is to observe that instead of seeing the
bare operators, $\sx{0}$, $\sz{0}\sz{1}$, $\sy{0}\sy{1}$, appearing
with varying prefactors, we see them appearing in the combinations
$h_0\sx{0},\,\frac{1}{\Omega^2}h_0^2J_{01}\sz{0}\sz{1},\,\frac{1}{\Omega^2}h_0J_{01}h_1\sy{0}\sy{1}$, 
where $\Omega$ is the energy scale at which the operator
appeared. There may be additional  prefactors, which we will
consider shortly, but first let us establish that these forms have an
invariant structure.

\subsection{Invariant Operators}

In this section we will show that the {\it form} of the 
operators defined as follows:
\be
\ba{c}
h_0\sx{0},\vspace{2mm}\\ 
h_0^2J_{01}\sz{0}\sz{1}\vspace{2mm}\\
h_0J_{01}h_1\sy{0}\sy{1}
\ea
\label{iop}
\ee
is preseved during the 
RG flow. The coefficients $h_0$, $h_0^2J_{01}$, and $h_0J_{01}h_1$
will only be replaced by their renormaized counterparts every time
that a decimation affects them. This is demonstrated in teh folloing examples.

The simplest example is the operator $h_0\sx{0}$. Under  the decimation of  $-J_{01}\sz{0}\sz{1}$, 
\begin{widetext}
\be
h_0\sx{0}\Rightarrow -\frac{h_{0}h_1}{J_{01}}\sy{0}\sy{1}=h^{eff}_{(01)}\sx{(01)}.
\ee
Another example is the operator
$\frac{1}{\Omega^2}h_0^2J_{01}\sz{0}\sz{1}$. The decimation of the
second bond, $-J_{12}\sz{1}\sz{2}$, modifies this operator as follows:
\be
h_0^2J_{01}\sz{0}\sz{1}\Rightarrow h_0^2J_{0(12)}\sz{0}\sz{(12)}
\ee
In the case of the operator $\frac{1}{\Omega^2}h_0J_{01}h_1\sy{0}\sy{1}$, the corresponding transformation $S_a=\frac{h_2}{2J_{12}}\sz{1}\sy{2}$ yields
\be
h_0J_{01}h_1\sy{0}\sy{1}\Rightarrow h_0J_{0(12)}\frac{h_1h_2}{J_{12}}\sy{0}\sx{1}\sy{2}=h_0J_{0(12)}h_{(12)}\sy{0}\sy{(12)}.
\ee

Decimating the second site in the chain, $-h_1\sx{1}$ involves the
transformation $S_a=\frac{J_{12}}{h_1}\sy{1}\sz{2}$, yielding the
following flow for the two operators $\frac{1}{\Omega^2}h_0^2J_{01}\sz{0}\sz{1},\,\frac{1}{\Omega^2}h_0J_{01}h_1\sy{0}\sy{1}$:
\be
h_0^2J_{01}\sz{0}\sz{1}\Rightarrow h_0^2\frac{J_{01}J_{12}}{h_1}\sz{0}\sx{1}\sz{2}=h_0^2J^{eff}_{02}\sz{0}\sz{2}.
\ee
The second transformation in this same decimation process is $S_b=\frac{J_{12}h_2}{h_1^2}\sz{1}\sy{2}$, which gives rise to:
\be
h_0J_{01}h_1\sy{0}\sy{1}\Rightarrow h_0\frac{J_{01}J_{12}}{h_1}h_1\frac{h_2}{h_1}\sy{0}\sx{1}\sy{2}=h_0J^{eff}_{02}h_2\sy{0}\sy{2}.
\ee
\end{widetext}
In all cases the flow due to decimations leaves the three forms of the
end operators invariant. We excluded here the cases of a decimation of
the first site or bonds; these are considered below. 

\subsection{Displacement Prefactors}

As mentioned before and seen from the results of (\ref{sxflow}), there
are still multiplicative prefactors coming before the invariant
operator forms. In Eq. (\ref{app2}), for instance, all three operators
from Eq. (\ref{iop}) have the prefactor
$\left(\frac{J_{01}}{h_0}\right)^2$. This suppression can be
associated with the {\it displacement} of the edge to the next
undecimated site. With the help of (\ref{sxflow})
it can be easily shown that an end site ($-h_0\sx{0}$) decimation
leads to the following flows:
\be
\ba{c}
h_0\sx{0}\Rightarrow \left(\frac{J_{01}}{h_0}\right)^2 h_{(01)}\sx{(01)}\vspace{2mm}\\
h_0^2J_{01}\sz{0}\sz{1}\Rightarrow\left(\frac{J_{01}}{h_0}\right)^2h_1J_{12}h_2\sy{1}\sy{2}\vspace{2mm}\\
h_0J_{01}h_1\sy{0}\sy{1}\Rightarrow\left(\frac{J_{01}}{h_0}\right)^2h_1^2J_{01}\sz{1}\sz{2}.
\ea
\ee
From the above equation we see that there is a factor $\left(\frac{J_{01}}{h_0}\right)^2$ associated with the displacement into the chain of the $\sz{0}\sz{1},\,\sy{0}\sy{1}$ operators. This  is repeated partially in the case of a bond decimation of the $-J_{01}\sz{0}\sz{1}$:
\be
\ba{c}
h_0^2J_{01}\sz{0}\sz{1}\Rightarrow h_{(01)}^2J_{(01)2}\sz{(01)}\sz{2}\vspace{2mm}\\
h_0J_{01}h_1\sy{0}\sy{1}\Rightarrow\left(\frac{h_0h_1}{J_{01}^2}\right)h_{(01)}J_{(01)2}h_2\sy{(01)}\sy{2}.
\ea
\ee

\subsection{Leading Order Correlations}

Now that we know how the end operators neglected in the text evolve
and discovered that their forms are invariant, we can show that these
operators do not change the leading order contributions to the
correlation functions of interest. 
The $xx$ correlation function is
\be
\ba{c}
C_L^{xx}=<h_0\sx{0}h_L\sx{L}>-<h_0\sx{0}><h_L\sx{L}>\vspace{2mm}\\
=<G|(h_0\sx{0}-<h_0\sx{0}>)(h_L\sx{L}-<h_L\sx{L}>)\pgs\vspace{2mm}\\
=\sum\limits_{\psi\neq G}^{}<G|h_0\sx{0}|\psi><\psi|h_L\sx{L}\pgs ,
\ea
\ee
with the sum over all excited states, $\psi$. 

As the RG process progresses, all the end operators will be generated
several times. However, we need only concern ourselves with the
last set of these generated. Previously generated edge operators will
have a larger suppression due to the more times they underwent edge displacement.

Considering the last decimation step which is needed to obtain the x-x
correlations and changing notation so that the remaining {\emph
  effective} sites are $\ell ,\,r$ as in the text, we have:
 \begin{widetext}
\be
\ba{c}
C_L^{xx}\approx \frac{e^{-\Lambda_{ell}-\Lambda_r}}{h_\ell h_r}\sum\limits_{\psi\neq G}^{}<H|h_{\ell }\sx{\ell }+\frac{1}{\Omega_{ell}^2}h_{\ell }^2J_{\ell r}\sz{\ell }\sz{r}+\frac{1}{\Omega_{ell}^2}h_{\ell }J_{\ell r}h_{r}\sy{\ell }\sy{r}|\psi>\vspace{2mm}\\
<\psi|h_{r}\sx{r}+\frac{1}{\Omega_r^2}h_{r}^2J_{\ell r}\sz{\ell }\sz{r}+\frac{1}{\Omega_r^2}h_{\ell }J_{\ell r}h_{r}\sy{\ell }\sy{r}\pni.
\ea	\label{summ}
\ee
\end{widetext}
The remaining low energy parts of the hamiltonian are
\be
\H=-h_\ell \sx{\ell }-J_{\ell r}\sz{\ell }\sz{r}-h_r\sx{r}. \label{fh}
\ee
Two cases need to be considered: a site decimation and a bond decimation. 
In the case of a bond decimation, the ground state of the system is
\be
|H^{(L)}>=\frac{1}{\sqrt{2}}(|\downarrow_{\ell}> |\downarrow_r>+|\uparrow_{\ell}> |\uparrow_r>)
\ee
and we immediately see that the only excitation that contributes in the sum (\ref{summ}) is the $\sx{\ell }\sx{r}$ term:
\be
C_L^{xx}\approx e^{-\Lambda_{\ell}-\Lambda_r}h_\ell h_r,
\ee
which shows that in this case the dangerous operators (Eq. \ref{dop})
do not contribute to the correlations.

The second case involves a site decimation. Let us assume that the dominant piece in (\ref{fh}) is $-h_\ell \sx{\ell }$:
\be
|H^{(L)}>=\frac{1}{2}(|\downarrow_{\ell}> +|\uparrow_{\ell}> )(|\downarrow_r>+|\uparrow_r>).
\ee
In this case the contribution of the $\sx{\ell }\sx{r}$ product is
only second order, and we need to consider the unitary transformation
that induces this decimation. To lowest order, this is
$S_a=-\frac{J_{\ell r}}{2h_\ell }\sy{\ell }\sz{r}$. Applying this to
(\ref{summ}) we get 
\begin{widetext}
\be
\ba{c}
C_L^{xx}e^{\Lambda_{\ell}+\Lambda_r}h_\ell h_r\vspace{2mm}\\\approx
\sum\limits_{\psi\neq G}^{}<H|h_{\ell }e^{iS_a}\sx{\ell }e^{-iS_a}+\frac{1}{\Omega_{\ell}^2}h_{\ell }^2J_{\ell r}\sz{\ell }\sz{r}+\frac{1}{\Omega_{\ell}^2}h_{\ell }J_{\ell r}h_{r}\sy{\ell }\sy{r}|\psi>\vspace{2mm}\\
<\psi|h_{r} e^{iS_a}\sx{r}e^{-iS_a}+\frac{1}{\Omega_r^2}h_{r}^2J_{\ell r}\sz{\ell }\sz{r}+\frac{1}{\Omega_r^2}h_{\ell }J_{\ell r}h_{r}\sy{\ell }\sy{r}\pni\vspace{2mm}\\
\approx\sum\limits_{\psi\neq G}<H<H|-h_{\ell }\frac{J_{\ell r}}{h_\ell }\sz{\ell }\sz{r}+\frac{1}{\Omega_{\ell}^2}h_{\ell }^2J_{\ell r}\sz{\ell }\sz{r}+\frac{1}{\Omega_{\ell}^2}h_{\ell }J_{\ell r}h_{r}\sy{\ell }\sy{r}|\psi>\vspace{2mm}\\
<\psi|h_{r}\frac{J_{\ell r}}{h_\ell }\sy{\ell }\sy{r}+\frac{1}{\Omega_r^2}h_{r}^2J_{\ell r}\sz{\ell }\sz{r}+\frac{1}{\Omega_r^2}h_{\ell }J_{\ell r}h_{r}\sy{\ell }\sy{r}\pni\vspace{2mm}\\
=h_\ell h_r\frac{J_{\ell r}}{h_\ell ^2}+\frac{(h_\ell -h_r)h_rJ_{\ell r}^2}{\Omega_r^2}-\frac{(h_\ell -h_r)h_rJ_{\ell r}^2}{\Omega_{\ell}^2}
\approx h_\ell h_r\frac{J_{\ell r}}{h_\ell ^2},
\ea
\ee
\end{widetext}
which is the same result as was derived in the text while ignoring the
additional edge operators from Eq. (\ref{dop}). In the above we used  $\frac{h_\ell ^2}{\Omega_{r,\ell}}\ll1$.

This demonstration can be repeated for the xx-z correlations and also
carried to higher order with the same conclusions. We have thus
verified that the energy correlations can be obtained from the leading
contributions to the flow of the edge energy operators in each step of
the RG.  

\bibliography{thebib}

\end{document}